\documentclass[]{aa}
\usepackage{graphicx,amssymb,amsmath}
\usepackage{txfonts}
\usepackage{lscape}
\usepackage{bm}
\usepackage{booktabs}
\usepackage{siunitx}   
\usepackage{paralist}
\usepackage{natbib}
\usepackage{hyperref}
\hypersetup{
    colorlinks=true,
    linkcolor=blue,
    citecolor=blue,
    urlcolor=blue
}

\newcommand{\st}{{\rm St}}
\newcommand{\ey}{{\bf \hat y}}
\newcommand{\ex}{{\bf \hat x}}
\newcommand{\ez}{{\bf \hat z}}
\newcommand{\eb}{{\bf \hat b}}
\newcommand{\ek}{{\bf \hat k}}
\newcommand{\kwave}{{\bf k}}
\newcommand{\en}{{\bf \hat n}}
\newcommand{\utotal}{{\bf  u}}
\newcommand{\ugas}{{\bf  v}}
\newcommand{\vdust}{{\bf  w}}
\newcommand{\bfield}{{\bf  B}}
\newcommand{\sbfield}{{\bf  b}}
\newcommand{\marray}{\it{\bf  M}}
\newcommand{\eigvect}{\it{\bf q}}
\newcommand{\cur}{{\bf  J}}
\newcommand{\ccoef}{{\cal C} }
\newcommand{\dcoef}{{\cal D} }
\newcommand{\Lambdaad}{\Lambda_{\rm AD} }
\newcommand{\etaad}{\eta_{\rm AD} }
\newcommand{\caz}{c_{A,z}}

\newcommand{\p}{\partial}

\newcommand{\bmB}{\bm{B}}

\newcommand{\rhog}{\rho_\mathrm{g}}

\begin{document}
\title{Streaming instabilities in weakly ionized protoplanetary discs: the Ambipolar Streaming Instability (AmSI)}
\titlerunning{Streaming instabilities in weakly ionized protoplanetary discs: The AmSI}
\author{Arnaud Pierens 
 \inst{1}
  \and
  Min-Kai Lin  \inst{2,3} }
\institute{ Laboratoire d'astrophysique de Bordeaux, Univ. Bordeaux, CNRS, B18N, all\'ee Geoffroy Saint-Hilaire, 33615 Pessac, France\\
\email{arnaud.pierens@u-bordeaux.fr}
\and
  Institute of Astronomy and Astrophysics, Academia Sinica, Taipei 10617, Taiwan
\and 
 Physics Division, National Center for Theoretical Sciences, Taipei City, 10617, Taiwan
  }

\abstract{
 The regions of protoplanetary discs where planets can form are believed to be weakly ionised, suggesting thereby that  non-ideal magneto-hydrodynamics (MHD) effects play an important role in the disc dynamics and in the planet formation process. In particular, the combined effect of  ohmic resistivity and  ambipolar diffusion can be responsible for launching MHD-driven disc winds.  In this context, we  focus on the effect of ambipolar diffusion (AD) and examine the stability of a dusty, magnetized disc by employing both  linear stability analyses and numerical simulations.  
 We show that  dust feedback tends to stabilize the MRI oblique modes involved in the ambipolar-shear instability.  We also find that ambipolar diffusion leads to the onset of a strong resonant drag instability (RDI),  in which an Alfv\'en wave is destabilized by the relative drift between the gas and dust components.  The main impact of AD is to modify the Alfv\'en wave frequency, resulting in a large resonance width. The instability is found to have significant growth rates even in  dust-poor discs  and for tightly coupled particles,  which may help to bridge the gap between  growth of dust grains through coagulation and planetesimal formation.   
}

\maketitle

\section{Introduction}
\label{sec:sec1}
The first stage of planet formation involves the coagulation of micron-sized grains into millimeter-sized dust \citep{dullemond05}. However, the bouncing \citep{zsom10} and fragmentation \citep{blum08} barriers tend to prevent further growth of solids, such that how particles grow beyond the millimeter-size range occurs remains an outstanding issue. The streaming instability \citep[hereafter SI,][]{youdin05,youdin07}, which arises from angular momentum exchange between the gas and dust components, has emerged as a leading mechanism to overcome these barriers. It has indeed been shown that gravitationally bound 100-1000 km-sized bodies can form directly through the gravitational collapse of the filaments produced during the non-linear evolution of the SI \citep{johansen09,simon16,abod19}. The original linear stability analysis, restricted to inviscid discs, indicates that the growth rate of the SI is the highest for solids marginally coupled to the gas, and for local dust-to-gas ratios $  \epsilon  $ higher than unity.

Recent works have extended this original linear analysis of the instability to examine whether the SI remains robust in the presence of external turbulence in the disc \citep{chen20,umurhan20}. In the particular case where turbulence is driven by the magneto-rotational instability \citep[hereafter MRI,][]{balbus91}, \citet{johansen07b} found that the SI can still operate and lead to the formation of structures that concentrate into high density regions; while \citet{balsara09} and \citet{tilley10} also reported significant dust clumping through the SI during the dust settling process. However, except in its innermost regions, where thermal ionization guarantees the MRI to operate, the bulk of the disc is expected to remain laminar due to a low ionization level of the gas. In these regions, non-ideal MHD effects play an important role in the disc dynamics, as these can drive magnetically driven disc winds, which are thought to be the dominant mechanism for angular momentum transport in these discs \citep{bai13,gressel15,bethune17}.

Regarding the effect of these non-ideal effects on the SI, \citet{yang18} considered a disc model sandwiched between a turbulent surface layer and a dead zone dominated by Ohmic resistivity. They found that strong clumping of solids can arise in the dead zone due to a weak radial diffusion of particles near the midplane. On the other hand, by examining the stability of a dusty, magnetized gas, \citet{lin22} found that in resistive discs, the SI can be stabilized by magnetic perturbations for small values of the dust-to-gas ratio. The consequence of the Hall effect on the SI was investigated by \citet{wu24}. By employing a combination of linear analysis and numerical simulations, these authors found a new instability resulting from the advection of magnetic perturbations relative to the dusty gas, which can dominate over the SI for small dust-to-gas ratios and weak magnetic fields.

In this paper, we focus on the effect of ambipolar diffusion on the SI. Given the importance of ambipolar diffusion in the context of wind-driven protoplanetary discs, this is an important issue to consider. We find evidence for a strong resonant drag instability \citep[hereafter RDI;][]{squire18a,squire18b,zhuravlev19} resulting from the destabilization of an ambipolar-modified Alfv'en wave by the relative drift between the gas and dust components. Compared to a standard Alfv\'{e}n wave RDI \citep{hopkins18a,lin22}, ambipolar diffusion allows the RDI to be triggered over a wider range of wavelengths, owing to its ability to modify the Alfv\'{e}n frequency. We also find that dust can stabilize the unstable oblique modes of the MRI \citep{kunz04,desch04} giving rise to the ambipolar-shear instability.

The paper is organized as follows. In Sect. \ref{sec:sec1}, we present the governing evolution equations of the gas and dust components. We then describe in Sect. \ref{sec:linear} the linear theory and present numerical solutions to the linear stability problem in Sect. \ref{sec:sec4}. In Sect. \ref{sec:sec5}, we discuss the effect of ambipolar diffusion on the Alfv'en wave RDI. In Sect. \ref{sec:sec6}, we present the results of nonlinear simulations of the instability. We finally summarise our findings in Sect. \ref{sec:sec7}.

\section{Physical model}
\label{sec:sec2}
\subsection{Shearing  sheet approximation}
We model a local patch in a protoplanetary disk within the shearing box framework \citep{goldreich65,latter17}. A cartesian coordinate system with origin located at an arbitrary distance $R_0$ from the central star corotates with angular velocity $\Omega=\Omega_k(R_0)$, with $\Omega_k$ the Keplerian angular velocity. The x- and y- axes are oriented radially outward and along the orbital direction respectively, while the z-axis is directed in the vertical direction. We assume that the domain is small compared to the orbital distance so that Keplerian rotation appears as a linear shear  flow with velocity ${\bf U_k}=-\frac{3}{2}x \Omega \ey$. We assume the system is axisymmetric and neglect the vertical component of stellar gravity. In these limits, the continuity and momentum equations for the gas and dust components are respectively given by:

\begin{equation}
\frac{\partial \rho_g}{\partial t}+\nabla \cdot (\rho_g \ugas)=0
\label{eq:rhog}
\end{equation}

\begin{equation}
\begin{split}
\frac{\partial \ugas}{\partial t}+\ugas \cdot \nabla  \ugas-2 v_y\Omega \ex +v_x\frac{\Omega}{2} \ey=2\eta R \Omega^2 \ex-\frac{1}{\rho_g}\nabla \left(P+\frac{B^2}{2\mu_0}\right)\\
+\frac{1}{\rho_g \mu_0}(\bfield\cdot\nabla)\bfield+\frac{\epsilon}{\tau_s}(\vdust-\ugas)
\label{eq:gasmom}
\end{split}
\end{equation}

\begin{equation}
\frac{\partial \rho_d}{\partial t}+\nabla \cdot (\rho_d \vdust)=0
\end{equation}

\begin{equation}
\frac{\partial \vdust}{\partial t}+\vdust \cdot \nabla  \vdust-2 w_y\Omega \ex +w_x\frac{\Omega}{2} \ey=-\frac{1}{\tau_s}(\vdust-\ugas)
\label{eq:vdust}
\end{equation}
 where $\rho_g$ and $\rho_d$ are the gas and dust densities respectively,  $\ugas$, $\vdust$ their velocities measured relatively to the background Keplerian shear, and $\mu_0$  the magnetic permeability. We adopt an isothermal equation of state with the pressure given by $P=\rho_g c_s^2$,  where $c_s=H_g \Omega$ is the (constant) sound speed with $H_g$  the gas pressure scale height. In Eq. \ref{eq:gasmom}, the term $\propto \eta$  corresponds to an outward force acting on the gas due to a background radial pressure gradient, and determined by the dimensionless parameter: 
 \begin{equation}
 \eta=-\frac{1}{2\rho_g \Omega^2 R_0}\frac{\partial P}{\partial r}
 \end{equation}
 Finally, $\epsilon=\rho_d/\rho_g$ is the dust-to-gas ratio and  $\tau_s$ is the stopping time that we will characterize in the following in terms of the dimensionless Stokes number $\st=\tau_s \Omega$.

The induction equation is given by: 
\begin{equation}
\frac{\partial \bfield}{\partial t}=\nabla\times (\ugas \times \bfield) - \frac{3}{2}\Omega B_x \ey + \nabla\times \left[ \frac{1}{\gamma_{in} \rho_i \rho_g} (\cur \times \bfield)\times \bfield\right],
\label{eq:ind}
\end{equation}
where $\cur\equiv \mu_0^{-1}\nabla\times\bfield$ is the current density, $\gamma_{in}$ is the ion-neutral drag coefficient and $\rho_i$ the ion mass density. 
\subsection{Equilibrium state}
In the following, we will  assume a constant magnetic field and null radial component. In practice, any radial component is sheared by differential rotation such that the radial component becomes rapidly much smaller that the azimuthal component. Under these conditions, there is no deviation from Keplerian rotation due to the magnefic field and the equilibrium conditions correspond to the traditional steady-state drift solutions for a dusty disc obtained by \citet{nakagawa86}. For constant values of $\rho_g$, $\rho_d$ and $\st$, equilibrium solutions to Eqs. \ref{eq:rhog}-\ref{eq:vdust} can be obtained,  leading to velocity deviations from Keplerian rotation given by: 
\begin{equation}
v_x=\frac{2\epsilon \st}{\Delta^2}\eta R_0\Omega,
\label{eq:uxgas}
\end{equation}
\begin{equation}
v_y=-\frac{(1+\epsilon+\st^2)}{\Delta^2}\eta R_0\Omega,
\end{equation}
\begin{equation}
w_x=\frac{-2 \st}{\Delta^2}\eta R_0\Omega,
\label{eq:vxdust}
\end{equation}
\begin{equation}
w_y=-\frac{(1+\epsilon)}{\Delta^2}\eta R_0\Omega,
\label{eq:uydust}
\end{equation}
where $\Delta^2=\st^2+(1+\epsilon)^2$. 

\section{Linear theory}
\label{sec:linear}
\subsection{Linearized equations}
In this section, we perturb Eqs. \ref{eq:rhog}-\ref{eq:ind} assuming a background field $\bfield=B_y\ey+B_z\ez$, with $B_y$ (resp. $B_z$) the azimuthal (resp. vertical) component of the magnetic field.  For any variable $A$, we also assume Eulerian perturbations such that:
\begin{equation}
A\rightarrow A+\delta A \exp [i(k_xx+k_zz)+\sigma t)]
\end{equation}
 where linearized quantities are indicated by the $\delta$ notation and where $k_x$ (resp. $k_z$) is the radial (resp. vertical) wavenumber.  Under these conditions,  the linearized equations for our dusty magnetized disc  are given by:
\begin{equation}
\sigma W=-i(k_x \delta v_x+k_z \delta v_z)-ik_xv_x W
\label{eqi}
\end{equation}
\begin{equation}
\begin{aligned}
\sigma \delta v_x=
&2\Omega \delta v_y-ik_xv_x \delta v_x-\frac{\epsilon}{\tau_s}(v_x-w_x)(Q-W)-\frac{\epsilon}{\tau_s}(\delta v_x-\delta w_x)\\
&-i k_x c_s^2W+ic_{A,z}^2(k_z \delta b_x-k_x\delta b_z-k_xB'_y\delta b_y)
\end{aligned}
\end{equation}
\begin{equation}
\begin{aligned}
\sigma \delta v_y=
&-\frac{\Omega}{2} \delta v_x-ik_xv_x \delta v_y-\frac{\epsilon}{\tau_s}(v_y-w_y)(Q-W)\\
&-\frac{\epsilon}{\tau_s}(\delta v_y-\delta w_y)+ic_{A,z}^2k_z \delta b_y
\end{aligned}
\end{equation}
\begin{equation}
\sigma \delta v_z=-ik_xv_x \delta v_z-\frac{\epsilon}{\tau_s}(\delta v_z-\delta w_z)-i k_z c_s^2W-ic_{A,z}^2k_z B'_y \delta b_y
\end{equation}
\begin{equation}
\sigma Q=-i(k_x \delta w_x+k_z \delta w_z)-ik_xw_x Q
\end{equation}
\begin{equation}
\sigma \delta w_x=2\Omega \delta w_y-ik_xw_x \delta w_x-\frac{1}{\tau_s}(\delta w_x-\delta v_x)
\end{equation}
\begin{equation}
\sigma \delta w_y=-\frac{\Omega}{2} \delta w_x-ik_xw_x \delta w_y-\frac{1}{\tau_s}(\delta w_y-\delta v_y)
\end{equation}
\begin{equation}
\sigma \delta w_z=-ik_xw_x \delta w_z-\frac{1}{\tau_s}(\delta w_z-\delta v_z)
\end{equation}
\begin{equation}
\sigma \delta b_x=ik_z  \delta v_x-i k_xv_x\delta b_x+\etaad(-k^2\delta b_x+k_xk_zB'_y \delta b_y)
\end{equation}
\begin{equation}
\begin{aligned}
\sigma \delta b_y=
&ik_z  \delta v_y-i(k_x\delta v_x+k_z\delta v_z)B'_y-ik_xv_x\delta b_y-\frac{3}{2}\Omega\delta b_x \\
&+\etaad\left[k^2\frac{k_x}{k_z}B_y^\prime\delta b_x-(k^2B_y^{\prime 2}+k_z^2)\delta b_y\right]
\end{aligned}
\end{equation}
\begin{equation}
\begin{aligned}
\sigma \delta b_z=
&-ik_x\delta v_x-ik_xv_x\delta b_z \\
&+\etaad(k_xk_z\delta b_x-k_x^2B'_y\delta b_y-k_x^2 \delta b_z)
\end{aligned}
\label{eqf}
\end{equation}
where $k=\sqrt{k_x^2+k_z^2}$,  $W=\delta \rho_g/\rho_g$, $Q=\delta \rho_d/\rho_d $, $\delta b_{x,y,z}=\delta B_{x,y,z}/B_z$, $B'_y=B_y/B_z$,  and where $\caz=B_z/\sqrt{\mu_0 \rho_g}$ is the Alfv\'en speed and $\etaad=\frac{B_z^2}{ \gamma_{in} \rho_i \rho_g}$ the ambipolar resistivity. The linearized system consisting of the previous equations corresponds to an eigenvalue problem:
\begin{equation}
\marray \eigvect=\sigma \eigvect
\label{eq:array}
\end{equation}
where $\marray$ is the matrix representation of the right-hand side of Eqs. \ref{eqi}-\ref{eqf} and $\eigvect$ is the eigenvector of  complex amplitudes. Here $\marray$ is a $11\times 11$ matrix and $\eigvect=[W,Q, \delta \ugas, \delta \vdust, \delta \sbfield]^T$.
We solve the stability problem numerically using the function \texttt{eig} of \texttt{NumPy}  to find the dimensionless growth rate $s=\sigma/\Omega$ as a function of the following parameters:
\begin{itemize}
\item $\st$: The Stokes number of the particle size
\item $\epsilon$: The dust-to-gas ratio
\item $\beta_z=\frac{c_s^2}{c_{A,z}^2}$: the plasma beta parameter
\item $\Lambdaad=\frac{\caz^2}{\etaad\Omega}$: the ambipolar Elsasser number
\item $B'_y$: the azimuthal field $B_y$ relative to the vertical field $B_z$.
\item $\tilde K_{x,z}=k_{x,z}\eta R$: the dimensionless  radial and vertical  wavenumbers
\end{itemize}

\subsection{Resonant Drag Instabilities}
It is generally expected that the relative drift between dust and gas can render a gas-dust mixture unstable to RDIs (\citet{squire18a,squire18b}). An RDI can be triggered whenever the gas can support waves that propagate at the same speed as the dust drift velocity. The  RDI condition is then  obtained by equating the frequency $\omega_{\rm gas}$ of the wave involved in the RDI to $\kwave\cdot(\vdust-\ugas)$, with $\kwave$ the wavevector. In the particular case of an  unstratified and axisymmetric shearing box,  the RDI condition then becomes:
\begin{equation}
\omega_{\rm gas}(k_x,k_z)=k_x(w_x-v_x).
\label{rdicondition}
\end{equation}
In this context, the classical streaming instability is an RDI driven by the destabilization of inertial waves  with frequencies given by:
\begin{equation}
\omega_{\rm gas}^2=\frac{k_z^2}{k_x^2+k_z^2}\Omega^2
\end{equation}
such that the RDI condition given by Eq. \ref{rdicondition} reads in that case:
\begin{equation}
\widetilde{K}_z^2=\frac{\widetilde{K}_x^2 \zeta_x^2}{1-\widetilde{K}_x^2\zeta_x^2}
\label{sirdi}
\end{equation}
with: 
\begin{equation}
\zeta_x^2=\frac{4\st^2(1+\epsilon)^2}{\Delta^4}\eta 
\end{equation}

In a magnetized disc, the gas also supports Alfv\'en waves that can be eventually destabilized by the dust-gas relative velocity when condition \ref{rdicondition} is fullfilled. Neglecting the effect of rotation for the time being, and assuming a purely vertical magnetic field, Alfv\'en waves have frequencies given by: 
\begin{equation}
\omega_{\rm gas}^2=k_z^2\caz^2,
\label{eq:alfvenrdi}
\end{equation}
resulting in an RDI condition given by: 
\begin{equation}
\widetilde{K}_z^2=\widetilde{K}_x^2 \zeta_x^2 \beta_z \frac{\eta}{h^2},
\label{alfvenrdi}
\end{equation}
where $h=H_g/R_0$ is the disc's aspect ratio. \citet{lin22} indeed found Alfv\'{e}n wave RDIs in dusty, magnetized discs, but they are easily suppressed by ohmic dissipation. However, we shall find that ambipolar diffusion can revive them in the presence of an azimuthal field. 
\\

\section{Numerical results}
\label{sec:sec4}

\begin{figure*}
\centering
\includegraphics[width=0.33\textwidth]{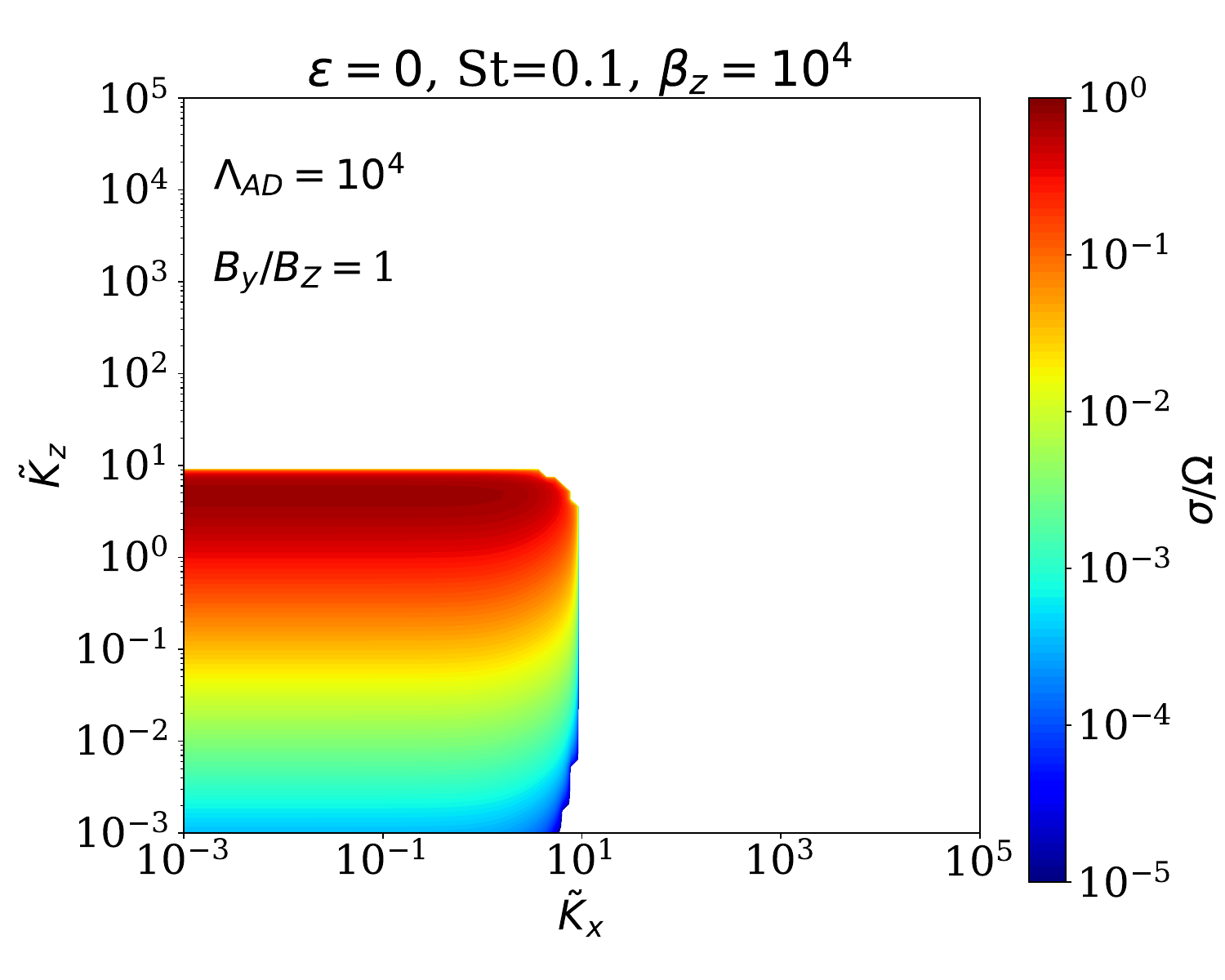}
\includegraphics[width=0.33\textwidth]{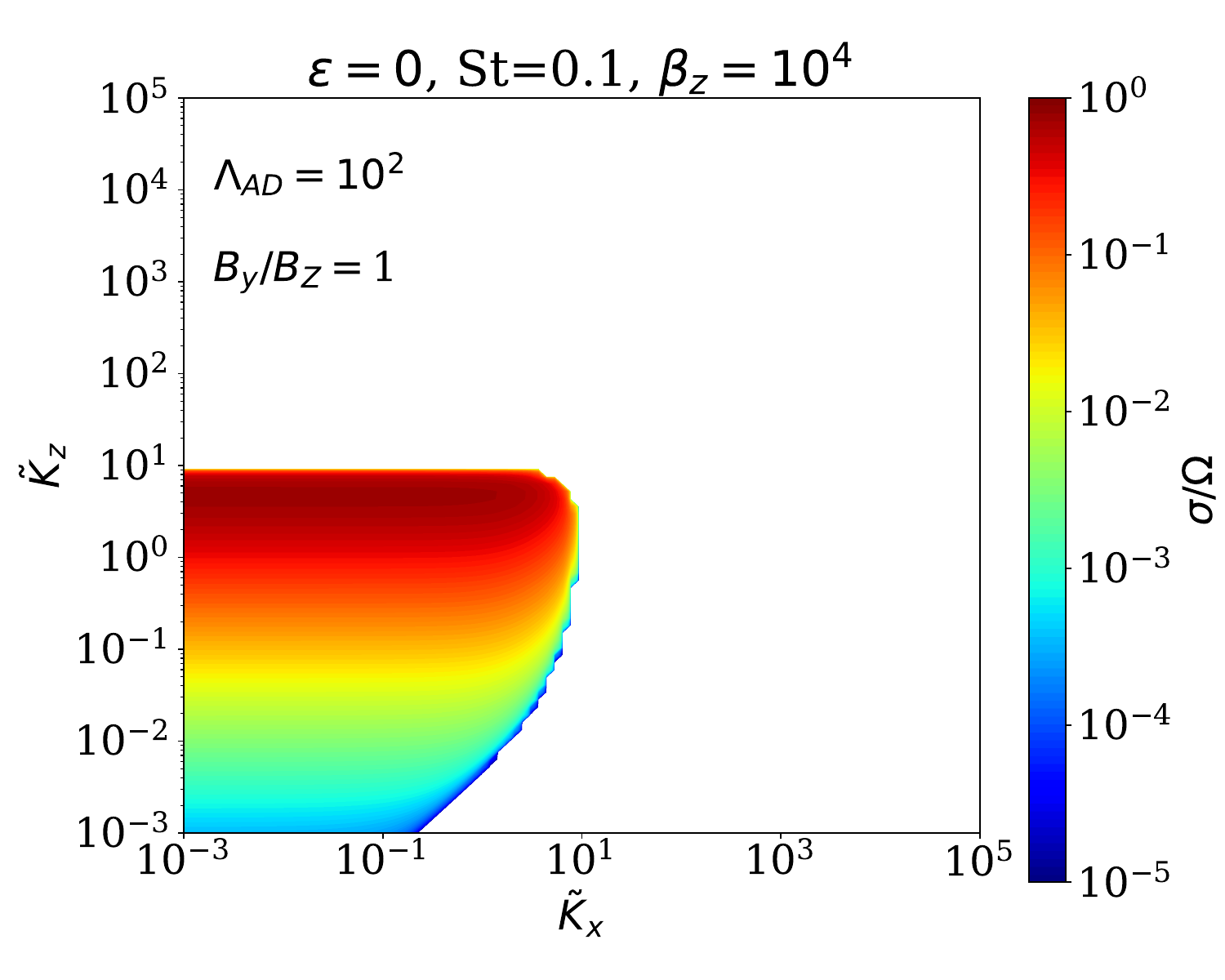}
\includegraphics[width=0.33\textwidth]{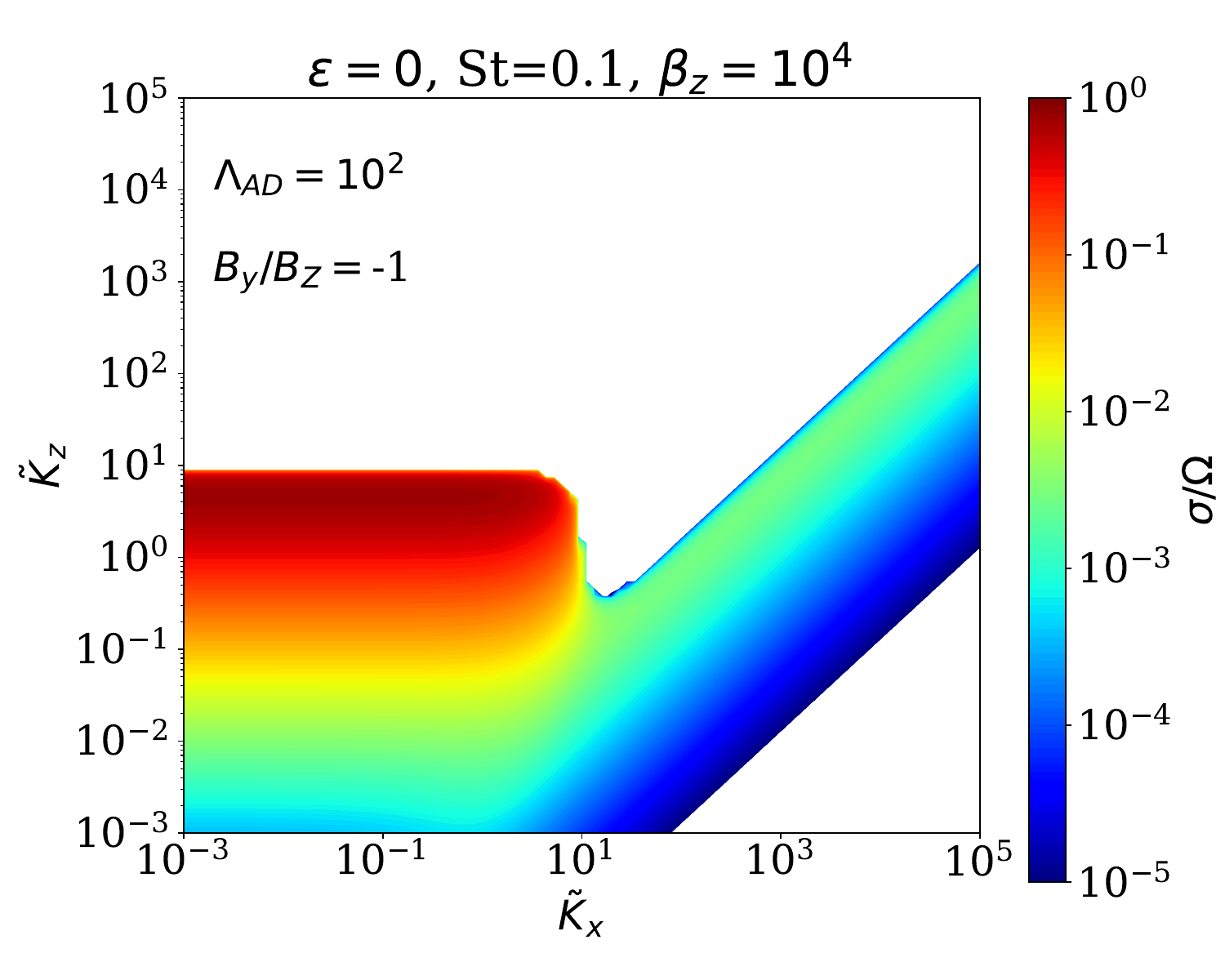}
\includegraphics[width=0.33\textwidth]{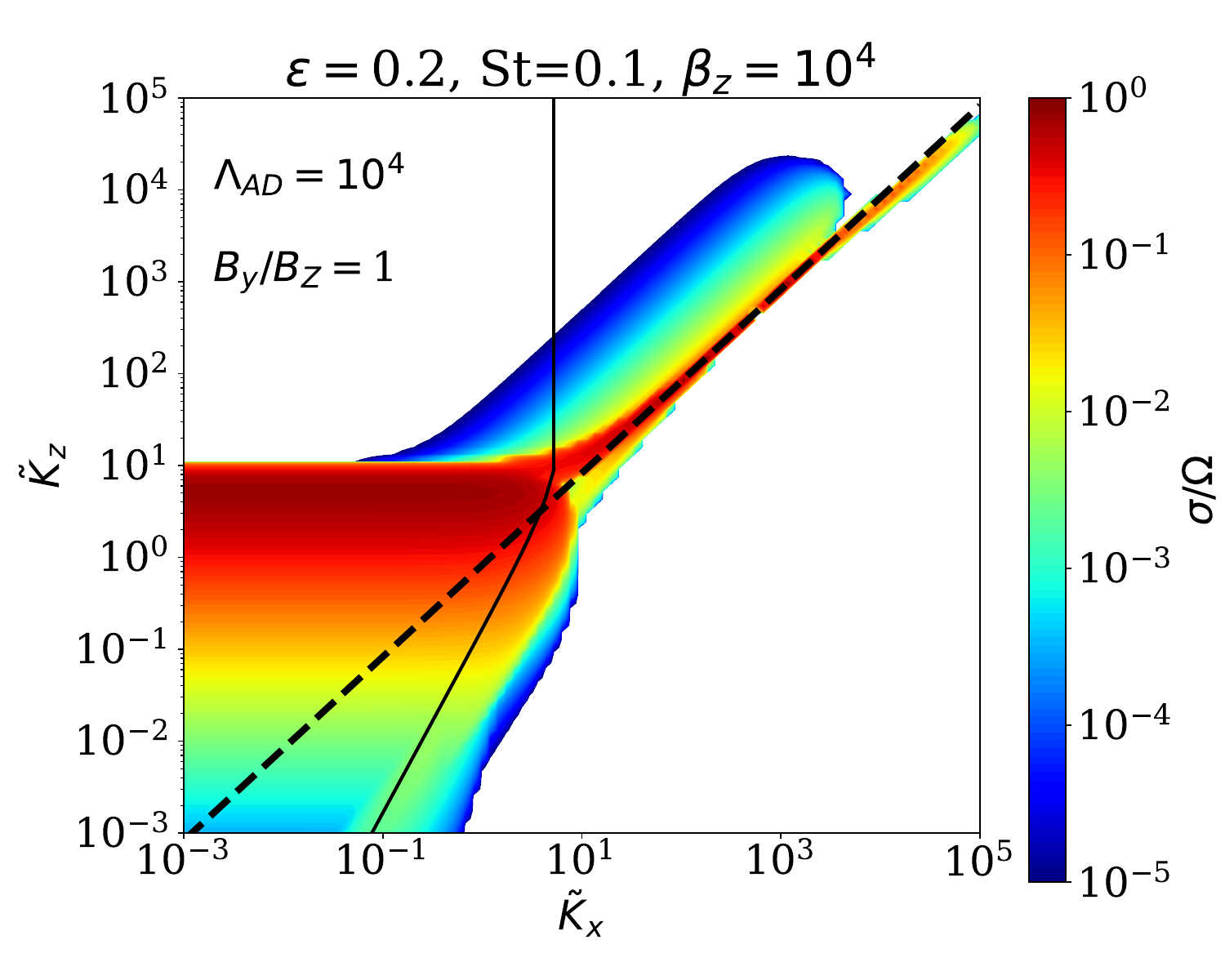}
\includegraphics[width=0.33\textwidth]{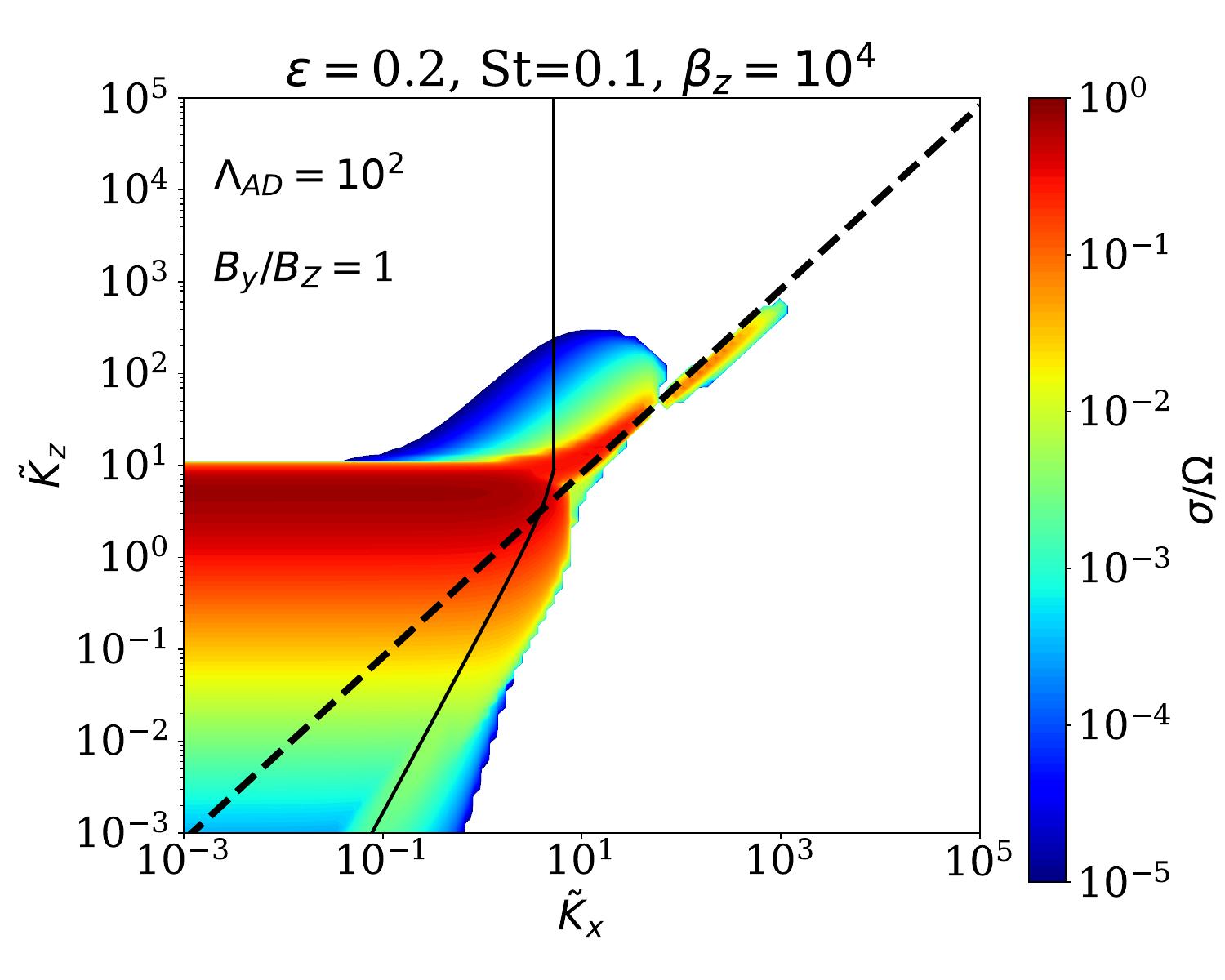}
\includegraphics[width=0.33\textwidth]{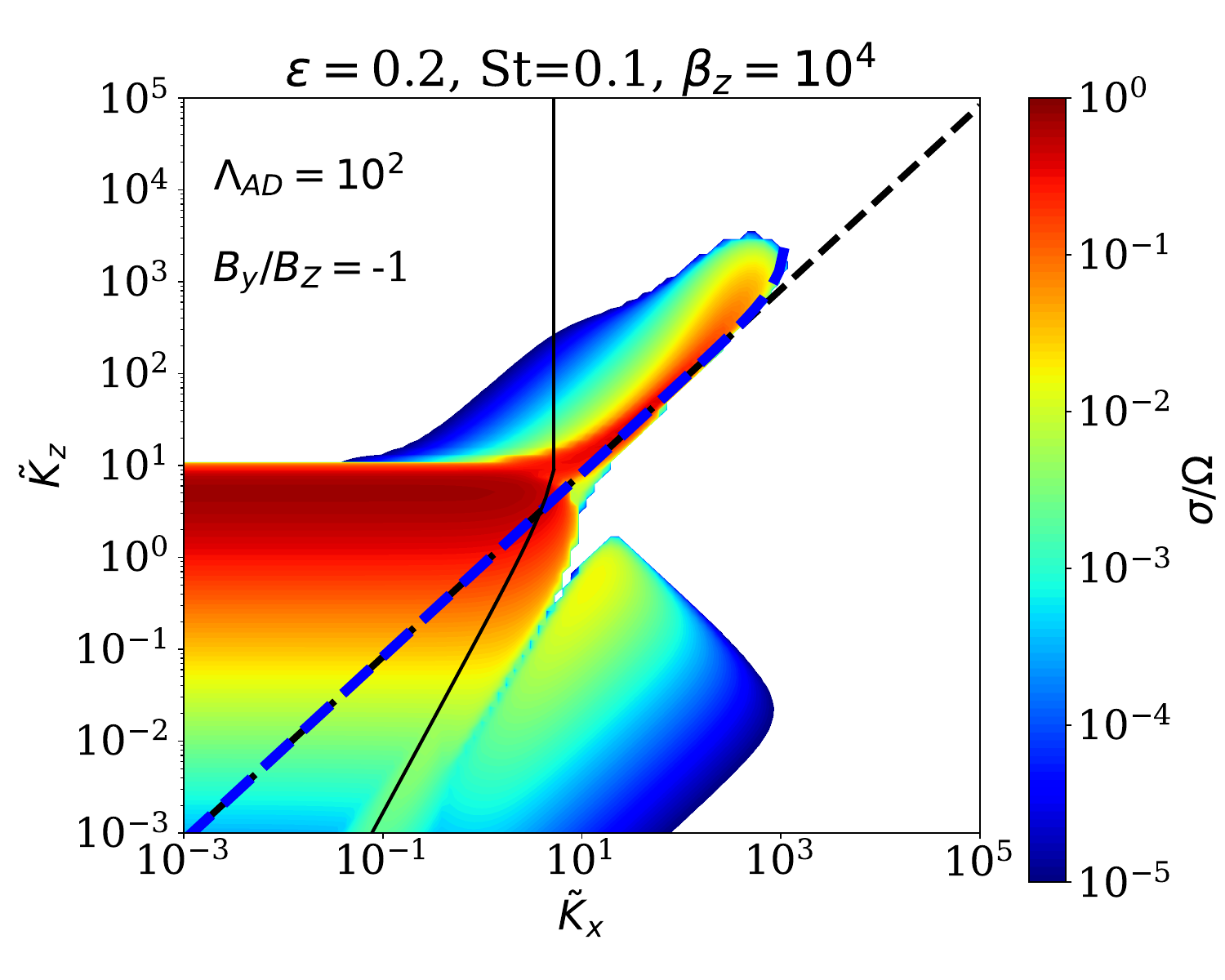}
\includegraphics[width=0.33\textwidth]{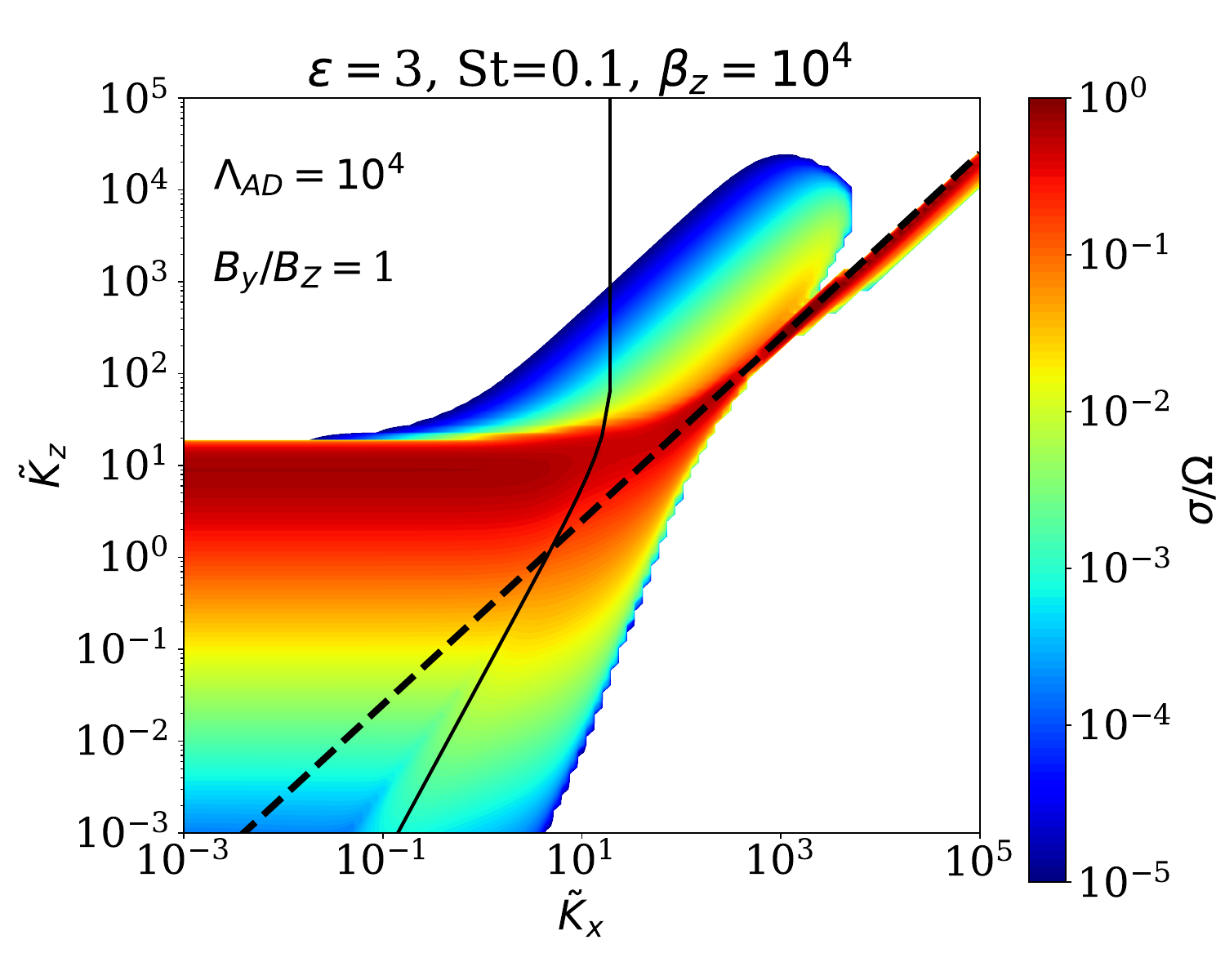}
\includegraphics[width=0.33\textwidth]{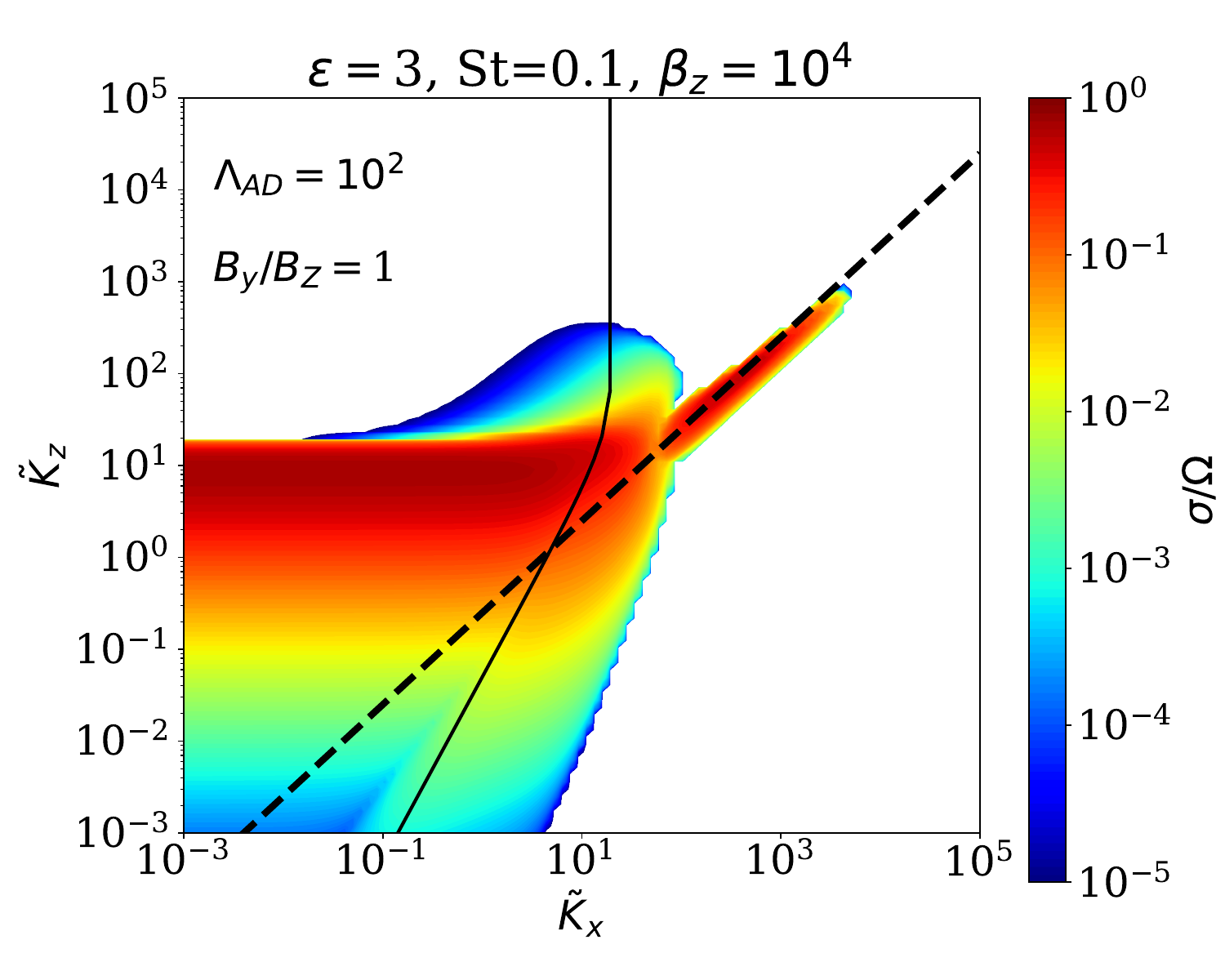}
\includegraphics[width=0.33\textwidth]{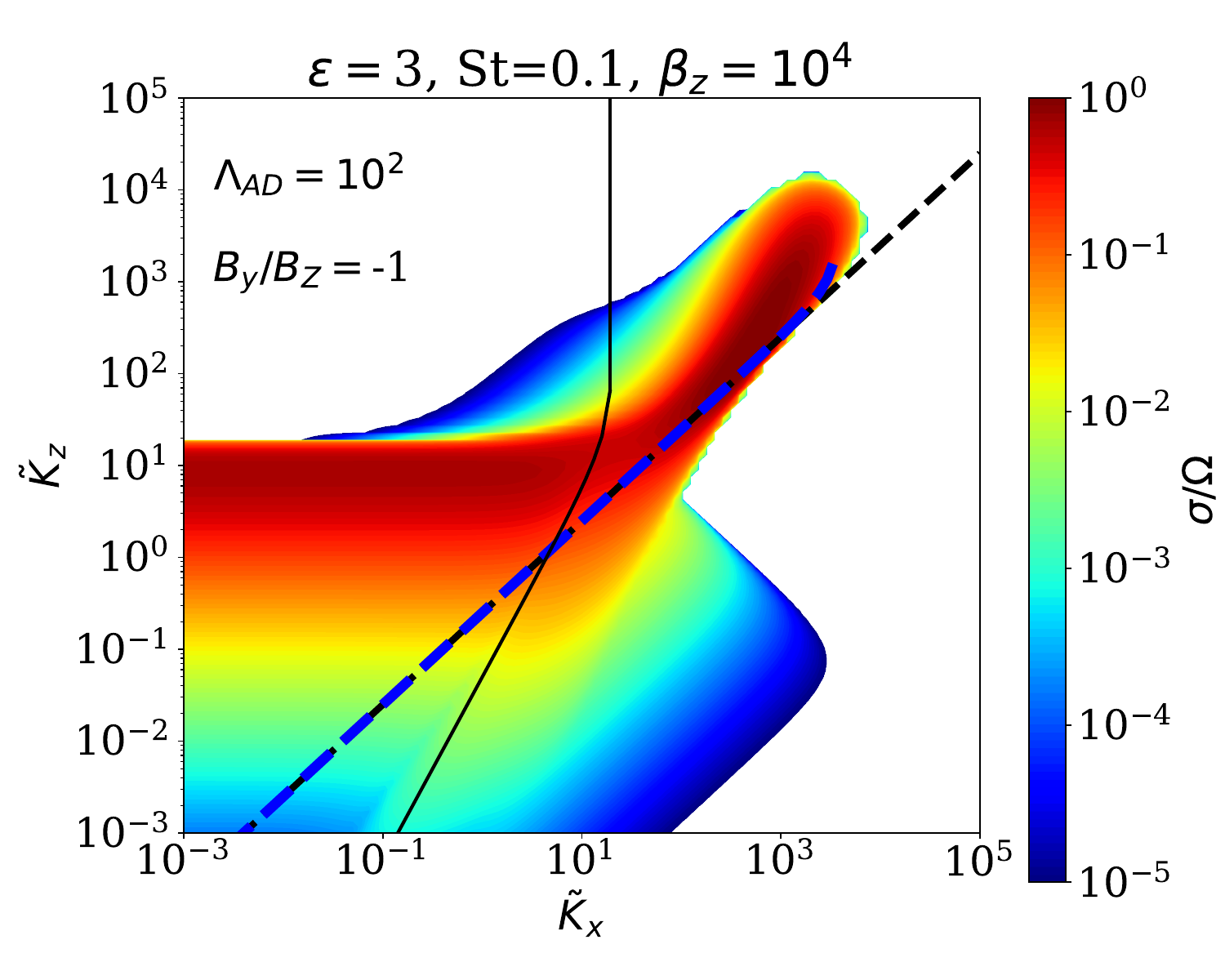}
\caption{Growth rates of unstable modes for different amplitudes of the azimuthal magnetic field,  for a dust-free disc with $\epsilon=0$ (top panel), a dust-poor disc with $\epsilon=0.2$ (middle panel),  a dust-rich disc with $\epsilon=3$. Here,   the Stokes number is fixed to  to $\st=0.1$ and the ambipolar Elsasser number to $\Lambdaad=100$.  The solid line  corresponds to the resonant drag instability (RDI) condition for the streaming instability (Eq. \ref{sirdi}), while the black dashed line corresponds to the RDI condition between Alfvén waves and the dust–gas radial drift (Eq.\ref{alfvenrdi}). The dashed blue line corresponds to the condition for AD-modified Alfv\'en wave RDI (Eq. \ref{eq:rdicondition}).}
\label{fig:reference}
\end{figure*}

\begin{figure*}
\centering
\includegraphics[width=\textwidth]{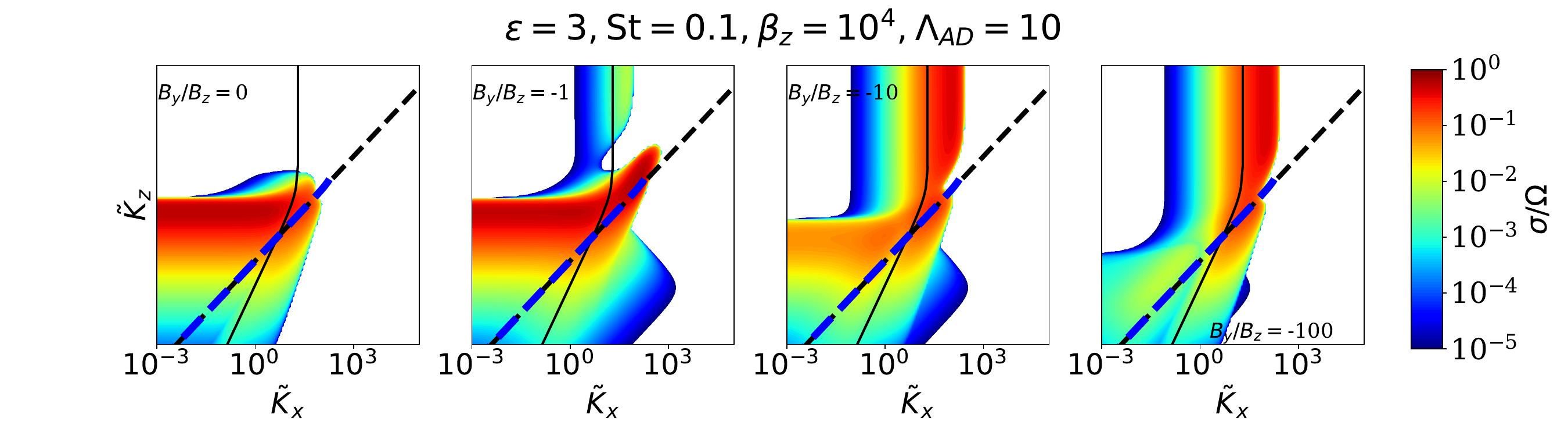}
\caption{Growth rates of unstable modes for different amplitudes of the azimuthal magnetic field. Here, the dust-to-gas ratio is fixed to $\epsilon=3$, the Stokes number to $\st=0.1$ and the ambipolar Elsasser number to $\Lambdaad=10$.}
\label{fig:varyby}
\end{figure*}

\begin{figure*}
\centering
\includegraphics[width=\textwidth]{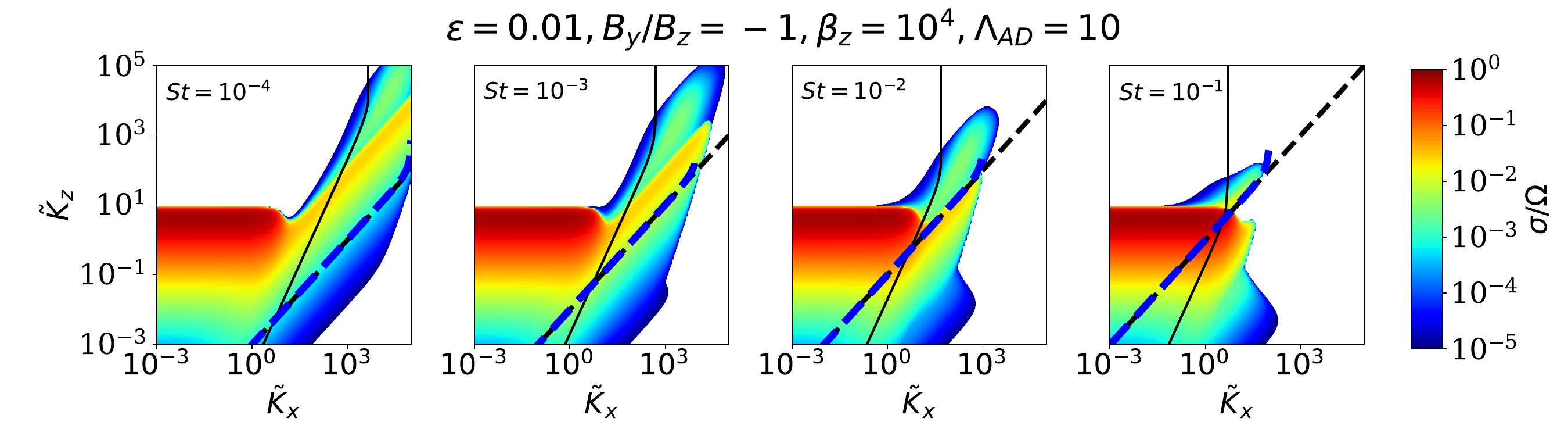}
\includegraphics[width=\textwidth]{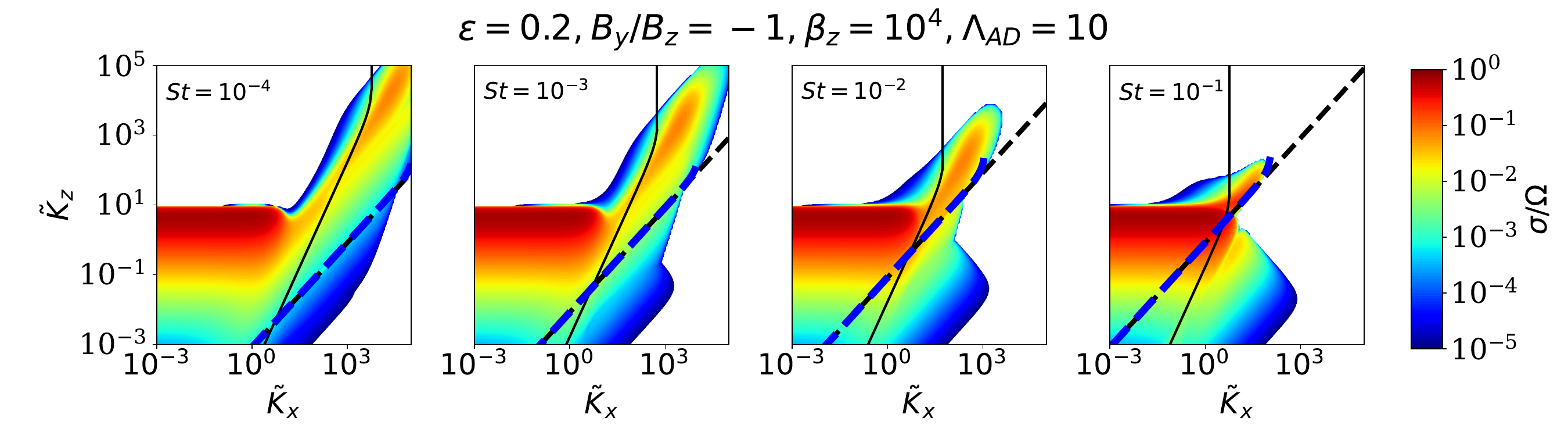}
\caption{Growth rates of unstable modes for different values of the Stokes number, for $\epsilon=0.01$ (top) and $\epsilon=0.2$ (bottom). Here,  the ambipolar Elsasser number to $\Lambdaad=10$ and the azimuthal field to $B_y/B_z=-1$.}
\label{fig:varyst}
\end{figure*}

\begin{figure}
\centering
\includegraphics[width=\columnwidth]{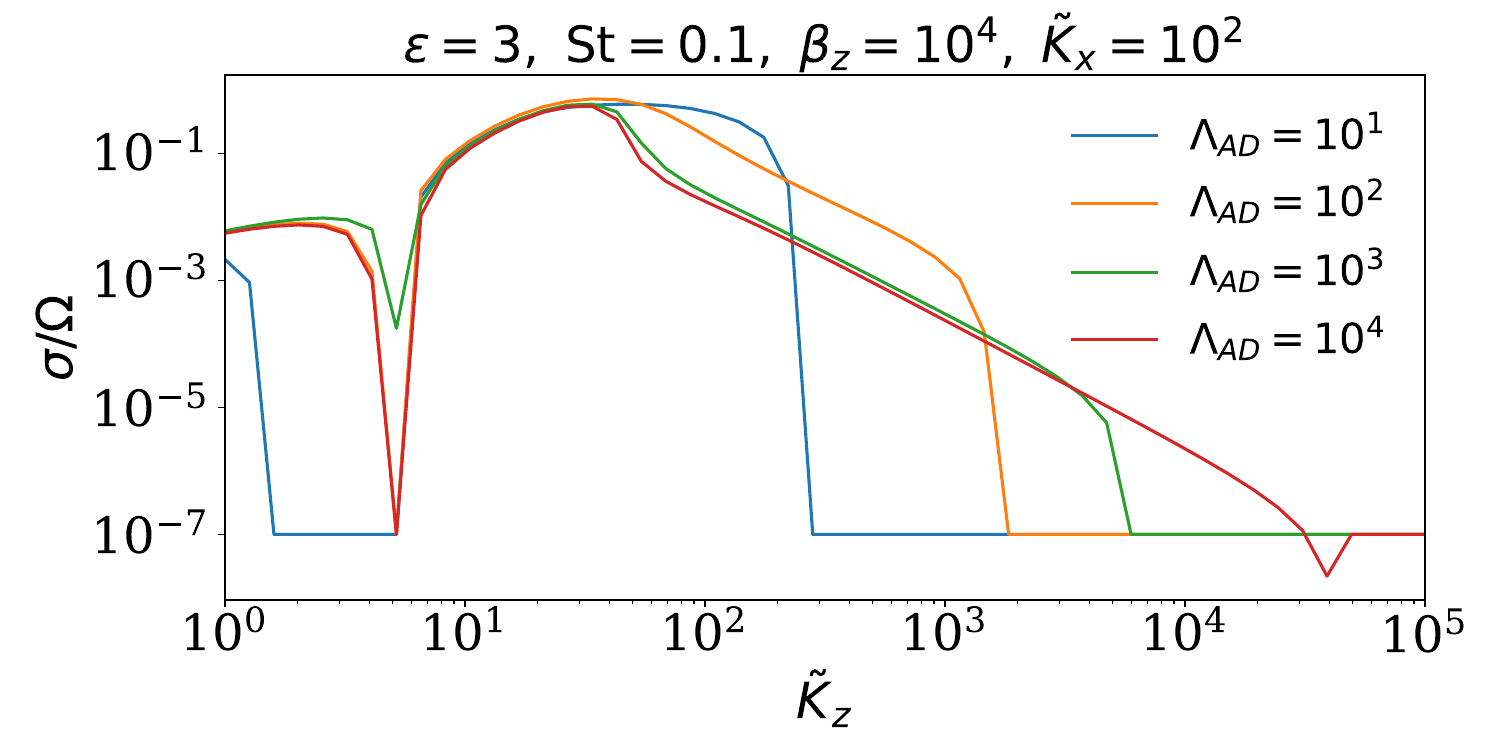}
\includegraphics[width=\columnwidth]{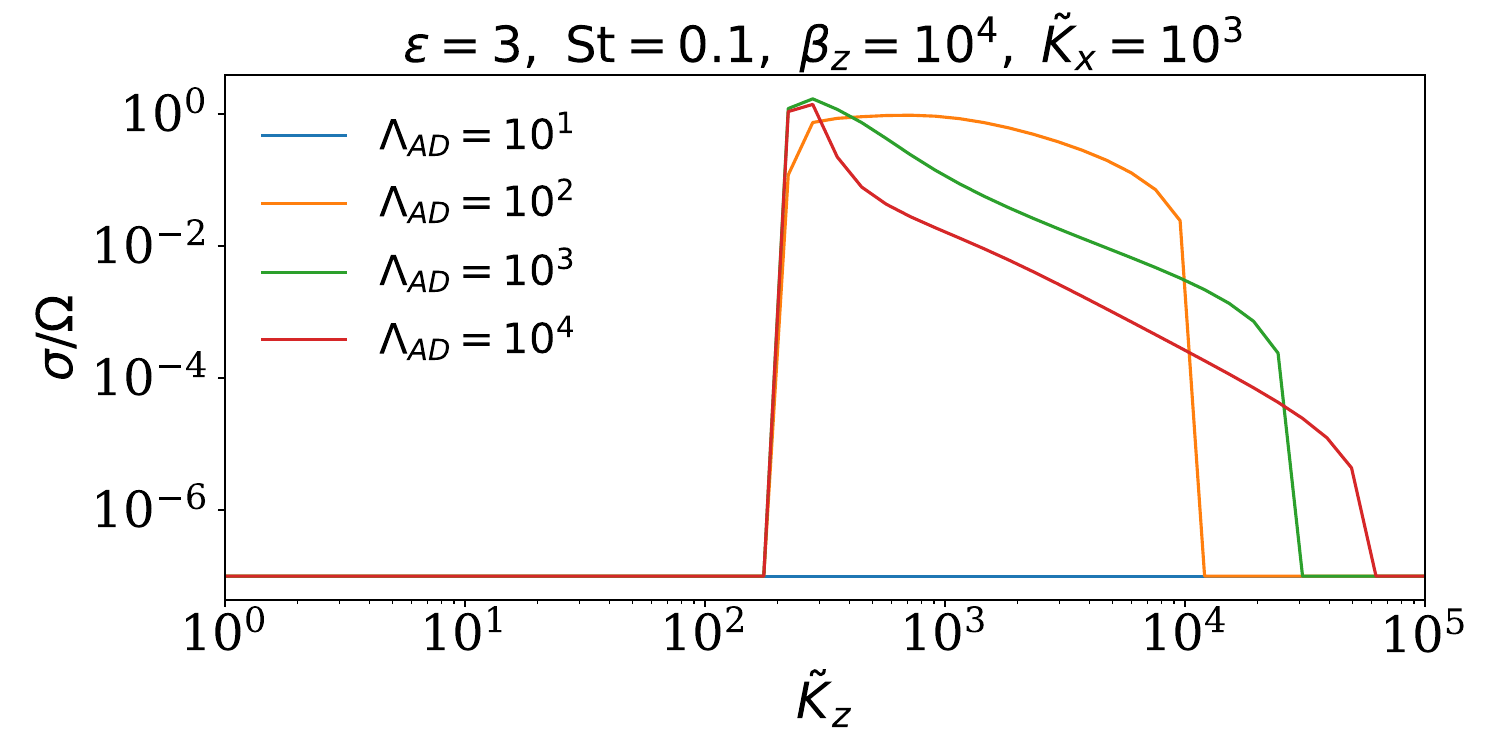}
\caption{Instability growth rate as a function of vertical wavenumber $\tilde K_z$ for different values of the ambipolar Elsasser number $\Lambdaad$,  for $\tilde K_x=10^3$ (upper panel) and $\tilde K_x=10^2$ (lower panel). Here, the dust-to-gas ratio is fixed to $\epsilon=3$ and the Stokes number to $\st=0.1$.}
\label{fig:kx2kx3.pdf}
\end{figure}

\begin{figure}
\centering
\includegraphics[width=\columnwidth]{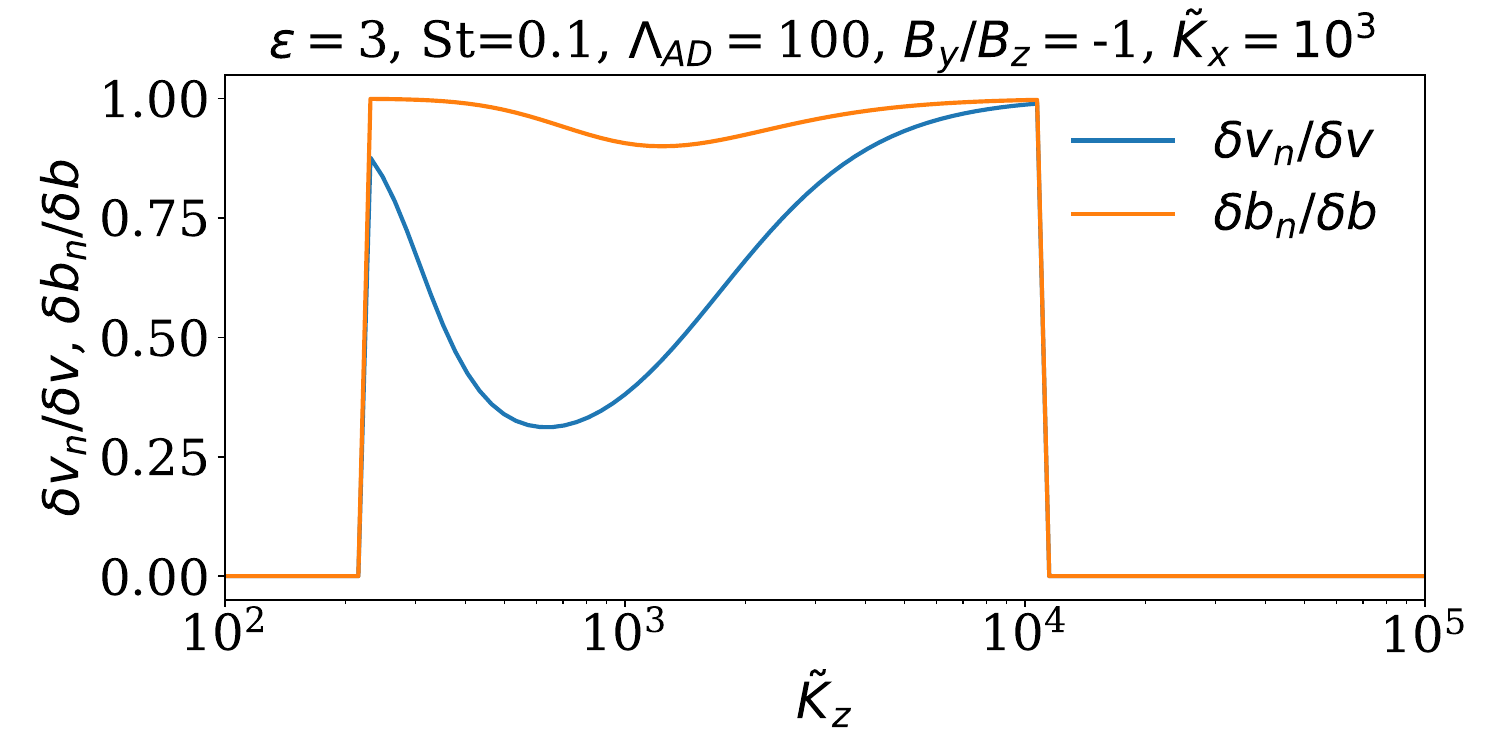}
\includegraphics[width=\columnwidth]{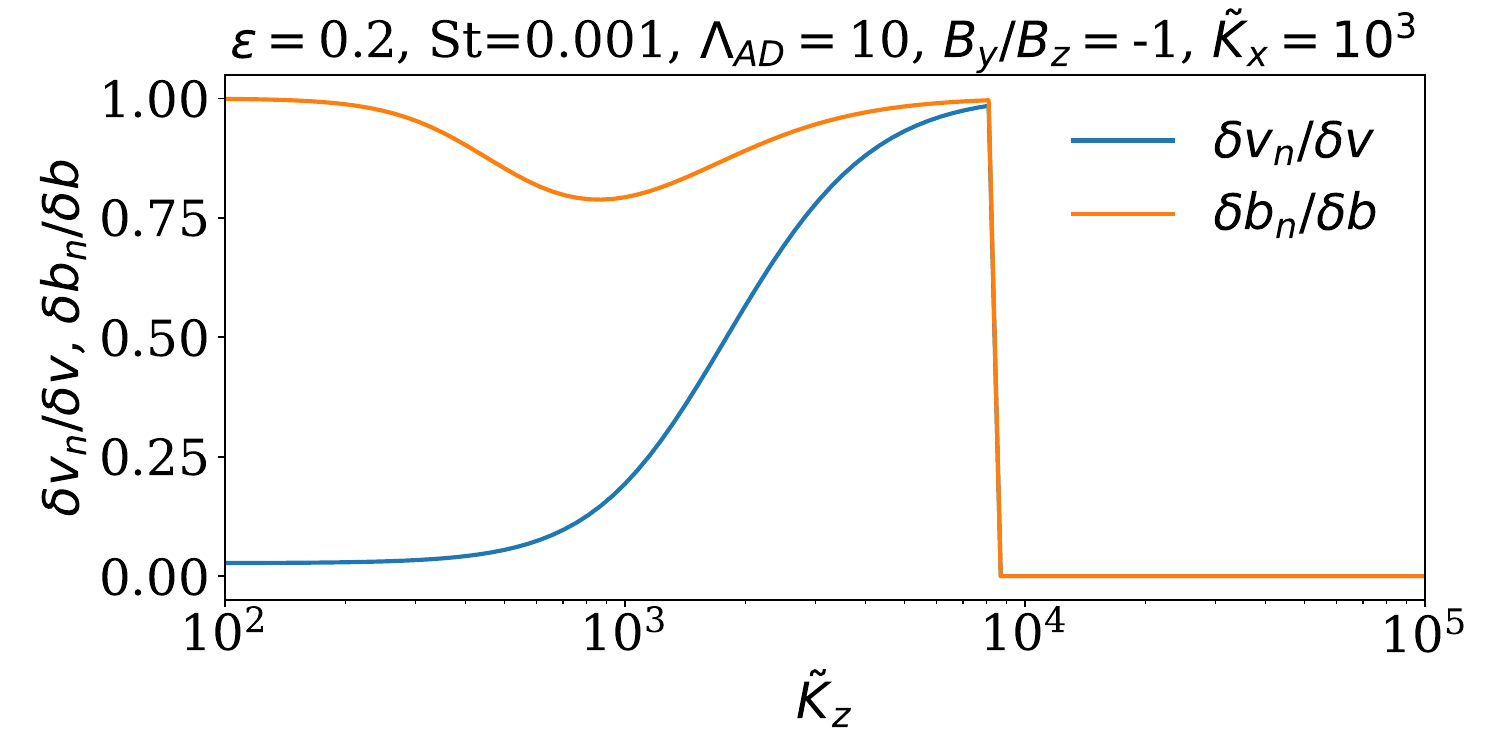}
\caption{Velocity and magnetic field components along $\en=\ek \times \eb$ as a function of vertical wavenumber $\tilde K_z$, for $\epsilon=3$, $\st=0.1$, $\Lambdaad=100$ (upper panel) and $\epsilon=0.2$, $\st=0.001$, $\Lambdaad=10$ (lower panel). Here, the azimuthal field is fixed to $B_y/B_z=-1$ and the radial wavenumber to $\tilde K_x=10^3$. }
\label{fig:vnbn}
\end{figure}

\begin{figure}
\centering
\includegraphics[width=\columnwidth]{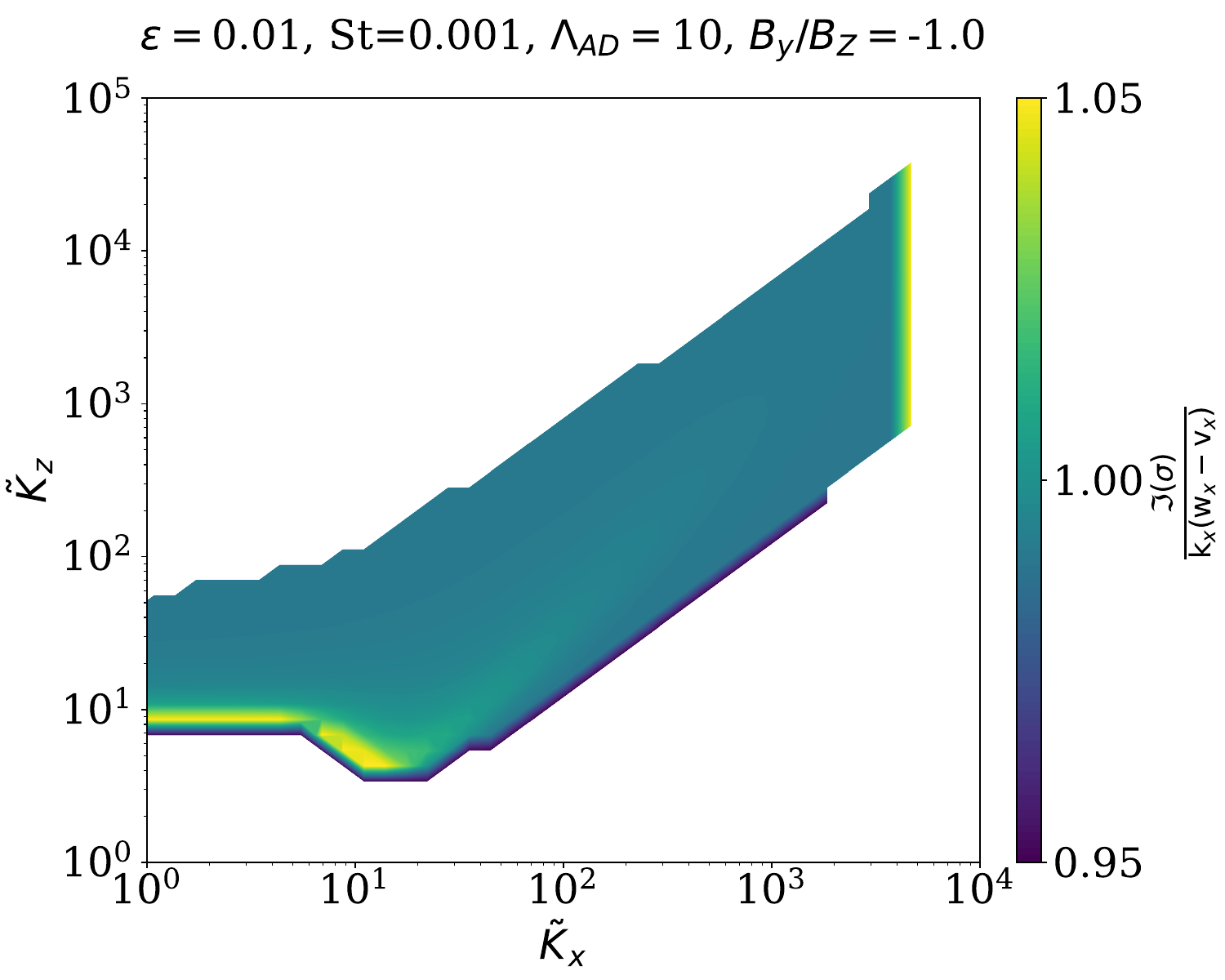}
\caption{For  $\epsilon=0.01$ and $\st=0.001$, ratio between oscillation frequencies  $\Im(\sigma)$ found by solving the stability problem (Eq. \ref{eq:array}) and frequencies expected from the resonance condition $k_x(w_x-v_x)$.}
\label{fig:checkrdi}
\end{figure}

\subsection{Effect of ambipolar diffusion on the SI}
In this section, we describe the numerical solutions to the aforementioned eigenvalue problem. To highlight the effect of ambipolar diffusion on the SI, we first vary $\epsilon$ and study the effect of changing the sign of $B_y$, while keeping the other parameters fixed. We then investigate the impact of changing $\st$ and the amplitude of the azimuthal field $B_y$.  

Fig. \ref{fig:reference} shows growth rates of unstable modes  as a function of wavenumbers, for $\st=0.1$, $\beta_z=10^4$ and for different values of $\epsilon$, that increases from upper to lower panels.  The left column corresponds to an almost ideal MHD configuration, with $\beta_z=10^4$, whereas the middle and right columns compare the cases $B'_y=1$ and $B'_y=-1$.  In this figure, the solid and dashed black lines represent the RDI conditions for the classical and  Alfv\'en wave SI \citep{squire18a,squire18b}, given respectively by Eqs. \ref{sirdi} and \ref{alfvenrdi}. Finally, the blue line shows the resonance condition for the AD-modified Alfv\'en wave RDI (see Sect. \ref{sec:rdi}). 

The upper row corresponds to the dust-free case. For $\Lambdaad=10^4$ (left panel), the unstable modes in the lower-left corner are the signature of the MRI, which should satisfy $k \caz<\sqrt{3} \Omega$  \citep{balbus91}. For $\Lambdaad=100$ and $B'_y>0$,  the impact of ambipolar diffusion is expected to be similar to that of ohmic resistivity, with a dissipation rate that decreases faster toward small wavenumbers \citep{sano99}. Compared to the ideal MHD case, adding ambipolar diffusion should in that case  stabilize modes with the highest wavenumbers, which is indeed what is observed. For $B'_y<0$,  oblique modes with $k_x\ne 0$ can be destabilized \citep{kunz04} by ambipolar diffusion, which couples in that case with differential rotation to amplify the magnetic field \citep{desch04}.  Since a value $\Lambdaad=100$ is considered here, the oblique modes have smaller growth rates than axial wavenumbers,  although  we note in passing  that they can  be more unstable than  the axial modes for $\Lambdaad<1$ \citep{lesur21}.

We now turn to the case of dusty magnetized discs.  The middle row of Fig. \ref{fig:reference} shows the growth rates for a dust-poor disc with $\epsilon=0.2$ and $\st=0.1$,  corresponding to the Lin B setup in the terminology of \citet{youdin07}.  In the ideal-MHD case (left panel),  the unstable modes clustering along the straight line for $\tilde K_x, \tilde K_z\gtrsim 10$ can be clearly identified as Alfv\'en wave streaming instabilities (AwSI) arising from the resonance between Alfv\'en waves and the dust-gas radial drift, since they follow the resonance condition given by Eq. \ref{eq:alfvenrdi}. For an AD-dominated disc with $\Lambdaad=100$ and $B'_y=1$ (middle middle panel), however, we find that these RDI unstable modes can be stabilized by AD. Again, as AD-dominated and resistive discs  are expected to behave similarly in that case, this is consistent with the results of \citet{lin22}  who found that the AwSI modes are easily damped by ohmic resistivity.  For $B'_y=-1$, we see that a primary effect of dust loading is to stabilize the MRI oblique modes of \citet{kunz04}. Moreover, compared to the $B'_y>0$ case,  a similar effect of AwSI mode damping is obtained for  $\tilde K_x, \tilde K_z \gtrsim 10^3$. For smaller wavenumbers with $10\lesssim \tilde K_x, \tilde K_z \lesssim 10^3$ , however,  we find that ambipolar diffusion, on the contrary, tends to increase the growth rates of the unstable modes located slightly away from the straight line corresponding to the resonance condition (Eq. \ref{eq:alfvenrdi}). Although ambipolar diffusion does not significantly increase the maximum growth rate compared to the ideal-MHD case, its main effect here is to extend the parameter space of unstable AwSI modes toward larger $\tilde K_z$. 

This  effect becomes more pronounced as $\epsilon$ increases. The bottom row of Fig. \ref{fig:reference} shows the growth rates for a dust-rich disc with $\epsilon=3$ and $\st=0.1$ \citep[corresponding to the Lin A setup in the terminology of][]{youdin07}. For $B'_y<0$ (bottom right panel),  the "flared" region of unstable modes with $10^2\lesssim \tilde K_x, \tilde K_z \lesssim 10^4$ arises from the effect of ambipolar diffusion. Again, AD does not increase the maximum growth rate of AwSI modes but rather increases the parameter space of unstable modes for which $\sigma \sim \Omega$. Compared to the ideal-MHD case, we see that ambipolar diffusion  indeed tends to destabilize modes with vertical wavelengths shorter by two orders of magnitude.  In the following, we will refer to the modes that are destabilized by AD as AmSI modes. We will show in Sect. \ref{sec:rdi} that these unstables modes are due to an RDI associated with a resonance between AD-modified Alfv\'en waves and the dust-gas radial drift.

\subsection{Parameter study}
In this section, we discuss how the the AD-driven instability described above depends on parameters, with a particular focus put on the effect varying the azimuthal field and the Stokes number. In Fig. \ref{fig:varyby}, we display growth rates for the Lin A setup ($\epsilon=3$, $\st=0.1)$ with $\Lambdaad=10$, and for different values of $B_y/B_z$. For $B'_y=-1$ (second panel), the AmSI modes appear as the band of unstable modes enclosed within the region delimited by $10^2 \lesssim \tilde K_x \lesssim 10^3$ and $10 \lesssim \tilde K_z \lesssim 10^2$.  Compared to the case with $\Lambdaad=10^2$ (bottom right panel in Fig. \ref{fig:reference}), the range of unstable modes is therefore truncated at larger wavenumbers,  due to a higher ambipolar  diffusivity.  The smaller Elsasser number  also causes the classic SI modes to appear in the blue-green rectangular region corresponding to  $1 \lesssim \tilde K_x \lesssim 10^2$ and $10^2 \lesssim \tilde K_z \lesssim 10^5$,  with growth rates that increase with the amplitude of the azimuthal field.  This is simply due to the fact that the ambipolar resistivity scales with $B^2$, such that increasing $|B'_y|$ pushes the system toward the hydrodynamic limit. 

Fig. \ref{fig:varyst} shows the growth rates for different values of the Stokes numbers, and for dust-poor disks with relatively small values $\epsilon=0.01, 0.2$.  For $\epsilon=0.01$ (upper panel), the growth rates remain relatively modest with $\sigma \approx 10^{-3}\Omega$ typically. This case, however,  i) confirms that dust can stabilize the original oblique modes of the MRI and shows  ii) that the AmSI modes seem to emerge from the stable branch of these oblique modes. Due to the small value for $\epsilon$, it seems unlikely that this is a consequence  of the reduction of Alfv\'en speed caused by dust loading. Although not shown here, we checked this by considering a pure magnetized gas disc, but employing an effective  Alfv\'en speed  given by:
\begin{equation}
c_{A,eff}=\frac{\caz}{\sqrt{1+\epsilon}}.
\end{equation}
For $\epsilon=0.2$ (lower panel), it is interesting to note that the AmSI modes can have growth rates of $\sigma\sim 0.1\Omega$, even for Stokes numbers as small as $\st=10^{-4}$. This suggests that,  in comparison with the classic SI modes, the conditions on $\epsilon$ and $\st$ for the unstable AmSI modes to develop with significant growth rates are less stringent. 

\section{Unstable AmSI modes}
\label{sec:sec5}

 We previously identified a new branch of unstable modes that we referred to as AmSI modes. The aim of this section is to understand the origin of the different properties of these AmSI modes that emerged from our numerical calculations. In the following, we discuss in particular how dust can stabilize the oblique modes of the ambipolar-shear instability. We also demonstrate that the AmSI modes can be interpreted as arising from an RDI driven by an AD-modified Alfv\'en wave and the relative drift between the gas and dust components. 

\subsection{Effect of dust on MRI oblique modes}
To investigate the effect of dust loading on the stability of the MRI oblique modes, we employ a single-fluid model of a magnetized, dusty gas disc, that we  present in details  in Appendix \ref{sec:appA}. We then make use of the dispersion relation that we obtain (see Eq. \ref{eq:disperse} ), and follow \citet{kunz04} by assuming that the transition from stability to instability occurs  through $\sigma \approx 0$. In the limit $\sigma\rightarrow 0$, Eq. \ref{eq:disperse} becomes:
\begin{equation}
\sigma= -\frac{\ccoef_0}{\ccoef_1}\left(1+\frac{\Re(\dcoef_1) \tau_s}{\ccoef_0}\right)
\label{eq:sigmawithdust}
\end{equation}
where the coefficients $\ccoef_0$  and $\ccoef_1>0$ are given in \citet[][see their Eqs. 34-35]{kunz04}, and where $\Re(\dcoef_1)$ is obtained from Eq. \ref{eq:d1} and reads:
\begin{equation}
\Re(\dcoef_1)=-2f_d\frac{B_y}{B_z}\frac{k_x}{k_z}(\caz k_z)^4\Omega+\frac{B_z^2}{\gamma \rho_i \rho_g}f_d \frac{B_y^2}{B_z^2}\caz^2 k_x^2 k_z^2 f_d\Omega^2 
\end{equation}

 where $f_d=\epsilon/(1+\epsilon)$.  The latter expression can also  be written as: 
\begin{equation}
\Re(\dcoef_1)=-2f_d\frac{B_y}{B_z}\frac{k_x}{k_z}(\caz k_z)^4\Omega\left[1-\frac{\Lambdaad^{-1}}{2}\frac{B_y}{B_z}\frac{k_x}{k_z}\right]
\label{eq:redcoef1}
\end{equation}
We note that,  in deriving Eq. \ref{eq:sigmawithdust}, $\dcoef_0$ was not taken into account as it is purely imaginary. For a pure magnetized gas disc, MRI oblique modes are unstable for  $\ccoef_0<0$ \citep{kunz04}, which requires the necessary condition $B_y k_x <0$ to be satisfied \citep{lesur21}. In that case, it is clear that $\Re(\dcoef_1)>0$  which leads, from Eq. \ref{eq:sigmawithdust}, to a smaller growth rate,  or even to a stabilization of the MRI oblique modes if $\Re(\dcoef_1)$ is large enough. This is fully  consistent with our numerical results, which show that adding dust indeed tends to stabilize the oblique modes with high wavenumbers (see, for instance, the middle-right panel of Fig. \ref{fig:reference}). Physically, this can be interpreted as resulting from the dust's back-reaction on the gas. It is indeed expected that it leads to a decrease in the  azimuthal velocity of the gas, such that the effect of the feedback loop between ambipolar diffusion and differential rotation is weakened. 

 Similarly,  for $\ccoef_0>0$ (but still assuming $B_y k_x <0$) , Eq. \ref{eq:sigmawithdust} confirms  that  dust  has a stabilizing effect, since the term in parentheses is positive. This suggests that, for  small Stokes numbers and dust-to-gas ratios,  the AmSI modes that seem to  originate from the  stable branch adjacent to the unstable oblique modes (e.g. upper-left panel of Fig. \ref{fig:varyst}) are not caused by the  effect of dust on the ambipolar-shear instability. 
\subsection{AD-modified Alfv\'en wave RDI: no rotation}
\label{sec:rdi}
The results presented above show also a clear trend for the instability region to lie along the RDI condition for the streaming instability of Alfven waves. To examine to which extent ambipolar diffusion affects the ability of Alfven waves to resonate with the dust-gas drift, we first  simplify the  analysis presented in Sect. \ref{sec:linear} by  neglecting rotation, and also consider the incompressible limit.  We will discuss in more details the effect of rotation in the next section.  Under these approximations, it is straightforward to show that the perturbed gas momentum and induction equations, projected along the direction of $\en=\ek \times \eb$, where $\ek$ and $\eb$ are unit vectors along the directions of the $\bf k$ and $\bfield$ respectively,   leads to the following dispersion relation for the magnetic perturbations oriented along $\en$:
\begin{equation}
\omega^2+i\omega \gamma B^2 k^2 \cos^2 \theta - k_z^2\caz^2=0
\label{eq:dispers}
\end{equation}
where $\gamma=1/\gamma_{in} \rho_i \rho_g$,  $\theta$ is the angle between $\ek$ and $\eb$ and with $\sigma=-i\omega $; while magnetic perturbations oriented along $\eb$ obey the following relation dispersion:
\begin{equation}
\omega^2+i\omega  \gamma B^2  k^2   - k_z^2\caz^2=0
\end{equation}
These expressions are similar to those derived by \cite{desch04} and \cite{pandey08}. 

In the following, we restrict to the case of magnetic perturbations directed along $\en$, which correspond to classical Alfven waves, as we show below that the instability described above can be interpreted as an RDI triggered when an AD-modified Alfven wave resonates with dust. In particular, this is supported by  Fig. \ref{fig:kx2kx3.pdf} which shows, for the case $\epsilon=3$, $\st=0.1$, $B_y/B_z=-1$, the instability growth rate as a function of $\tilde K_z$ for different values of $\Lambdaad$ (bottom right panel in Fig. \ref{fig:reference}), and for $\tilde K_x=10^2, 10^3$. It is clear that as $\Lambdaad$ increases, i.e. as the ideal MHD regime is approached,  the range of vertical modes unstable to pure Alfven wave streaming instabilities is recovered (bottom left panel in Fig. \ref{fig:reference}). 

To  clearly demonstrate that magnetic perturbations oriented along $\en$ may be involved in the instability, we plot in Fig. \ref{fig:vnbn} the velocity and magnetic field components along this direction as a function of vertical wavenumber $\tilde K_z$, for $\epsilon=3$, $\st=0.1$, $\Lambdaad=100$ (corresponding to the second panel in Fig. \ref{fig:varyby}) and $\epsilon=0.2$, $\st=0.001$, $\Lambdaad=10$ (corresponding to the lower panel of Fig. \ref{fig:varyst}). We see that for the range of unstable vertical modes with the highest growth rates, the velocity and magnetic perturbations are essentially directed along $\en$, which supports the transverse wave interpretation whose dispersion relation is given by Eq. \ref{eq:dispers}.

We note that this relation dispersion can also be written as: 
\begin{equation}
\omega^2+i\omega \etaad k_z^2  - k_z^2\caz^2=0
\label{eq:dispers2}
\end{equation}
Roots of Eq. \ref{eq:dispers2} have real and imaginary parts respectively given  by:

\begin{equation}
\Re(\omega)=\caz k_z\left(1-\frac{\etaad^2k_z^2}{4\caz^2}\right)^{1/2}
\label{eq:rew}
\end{equation}

and,

\begin{equation}
\Im(\omega)=-i\frac{\etaad k_z^2}{2}
\end{equation}

Solving for the resonance condition $\Re(\omega)=k_x(w_x-v_x)$, where $\Re(\omega)$ is given by Eq. \ref{eq:rew} and with $v_x$, $w_x$ given by Eqs. \ref{eq:uxgas} and \ref{eq:vxdust} respectively, leads to the following relation between the vertical and radial dimensionless wavenumbers:

\begin{align}
   \widetilde{K}_z^2 = \frac{2\Lambda_{\mathrm{AD}}^2\eta^2\beta_z}{h^2}\left(1\pm\sqrt{1-\frac{\widetilde{K}_x^2\zeta_x^2}{\Lambda_{\mathrm{AD}}^2\eta}}\right) 
\label{eq:rdicondition}   
\end{align}

In the limit $D\equiv \frac{\widetilde{K}_x^2\zeta_x^2}{\Lambda_{\mathrm{AD}}^2\eta}\ll 1$, we can expand the square root to $O(D^2)$ to obtain, by taking the negative solution,
\begin{equation}
\widetilde{K}_z^2 \approx \frac{
\eta \zeta_x^2 \beta_z \widetilde{K}_x^2}{h^2}\left(1+\frac{\zeta_x^2 \widetilde{K}_x^2}{4\eta \Lambdaad^2}\right),
\end{equation}
then Eq. \ref{alfvenrdi} is readily recovered in the limit of a vanishing AD resistivity. 

The resonance condition given by Eq. \ref{eq:rdicondition} is marked by the blue line in Figs. \ref{fig:reference}, \ref{fig:varyby} and \ref{fig:varyst}. For $\st=0.1$ (Fig. \ref{fig:reference}), we see that the model reproduces  the branch of unstable modes that appear in the $B_y<0$ case reasonably well.  For small Stokes numbers (Fig. \ref{fig:varyst}), however, the agreement is less clear, as the branch of unstable modes  emerging from the original MRI oblique modes does not closely follow the resonance condition. To check whether or not an RDI really operates in that case, we compared for each $( k_x,  k_z)$ the oscillation frequencies found numerically  by solving the stability problem represented by Eq. \ref{eq:array}, i.e. corresponding to $\Im(\sigma)$, to the frequency expected from the resonance condition $k_x(w_x-v_x)$.  The results of this procedure, for $\epsilon=0.01$ and $\st=0.001$, are displayed in Fig. \ref{fig:checkrdi}. It is clear that for each unstable $(k_x,k_z)$ pair, the two oscillation frequencies are comparable, which seems to support the RDI interpretation.

\subsection{Near-resonant growth}

Compared  with the ideal MHD case, the results presented above suggest that the primary effect of ambipolar diffusion is to widen the range of  unstable modes for which Alfv\'en waves can be involved in a RDI. To explain the significant width of the resonance that is obtained, we can notice that in presence of ambipolar diffusion, the system is almost identical to a forced oscillator with characteristic equation given by: 
\begin{equation}
\omega^2+i\omega \etaad k_z^2  - k_z^2\caz^2=\omega_{RDI}^2
\label{eq:oscillator}
\end{equation}
where we set $\omega_{RDI}=k_x(w_x-v_x)$, and whose oscillation amplitude $\mathcal{A}_{RDI}$ would be given by: 

\begin{equation}
\mathcal{A}_{RDI} \propto \frac{1}{\sqrt{(\omega_{RDI}^2-k_z^2\caz^2)^2+(\etaad k_z^2 \omega_{RDI})^2}}
\end{equation}
The equivalent quality factor associated with this forced oscillator, which is related to  the resonance width $\Delta \omega$  through: $ {\cal Q}=k_z \caz / \Delta \omega$, reads:
\begin{equation}
 {\cal Q}=\frac{\caz}{\etaad  k_z}=\frac{\Lambdaad}{\tilde K_z}\sqrt{\beta}h , 
 \end{equation}
where $h$ is the disc aspect ratio.
This corresponds to a resonant width given by:
\begin{equation}
\Delta \omega= \frac{\tilde K_z^2}{\Lambdaad \beta \eta} \Omega
\end{equation}
Setting $\Lambdaad=100$, $h=0.05$, we get ${\cal Q}=0.5$ for $\tilde K_z=10^3$, such that we indeed expect the  resonance to have a significant width in presence of ambipolar diffusion, 
In particular, for the parameters adopted here, $k_z \caz\approx 200 \Omega$ and the resonance width is estimated to $\Delta \omega \approx 400 \Omega$.

\subsection{Alfven-modified RDI: effect of rotation}

Although  the  instability we found  may be interpreted by the destabilization of an AD-modified Alfv\'en wave by the radial drift between dust and gas, it remains unclear why it disappears for $B_y k_x>0$. We suggest that, similar to the unstable oblique modes in non-ideal MHD, the instability also originates from the combination of shear and ambipolar diffusion, such that it can operate only under certain field geometries. In order to examine how the resonance condition given by Eq.  \ref{eq:rdicondition} is impacted by the effect of shear, we start by writing down the dispersion relation of pure MHD waves in the presence of rotation \citep{kunz08}:
\begin{equation}
\begin{aligned}
(\omega^2+i\omega \gamma B_z^2 k_z^2 - k_z^2\caz^2)&(\omega^2+i\omega \gamma B^2 k^2- k_z^2\caz^2)=\\
-2A\gamma (k_xB_y)(k_zB_z)\left[\omega^2-k_z^2 \caz^2\right]
\end{aligned}
\label{eq:wrotation}
\end{equation}
where $A=\partial \Omega / \partial \log R$. To extract the effect of rotation, we look for a correction to the AD-modified Alfv\'en wave frequency by writing $\omega=\omega_0+\omega_1$, where $\omega_0$ is given by Eq. $\ref{eq:rew}$ and with $\omega_1\propto A$. By  evaluating the first derivative of the left-hand side term  together with the right-hand side term of \ref{eq:wrotation} with respect to $\omega$ and evaluating it at $\omega=\omega_0$, and then equating terms of order $A$, we obtain:

\begin{equation}
\omega_1=2Ai\frac{ (k_x B'_y)\gamma B_z^2}{\caz}\frac{ k_z^2}{k_x^2+k^2B_y^{\prime 2}}
\end{equation}
 Given that the growth rate is $\sigma=-i \omega$ and $A<0$, we see that an azimuthal field with orientation such that $k_x B_y<0$ results in  $\sigma>0$, as expected. At the opposite,   $k_x B_y>0$ leads to a  damping of Alfv\'en waves through ambipolar diffusion.  For $k_x B_y>0$,  we can estimate under which conditions Alfv\'en waves are strongly damped by simply assuming $\omega_1=\omega_0$ in that case. We find that this occurs for: 

\begin{equation}
\Lambdaad<\frac{2\lvert A\lvert   \Omega^{-1} k_x k_z B'_y}{k_x^2+k^2B_y^{'2}}.
\end{equation}
For $k_x=k_z$ and $B'_y=1$, this leads to $\Lambdaad<1$, consistently with what is observed in Fig. \ref{fig:reference}.

\section{Direct simulations}
\label{sec:sec6}

We demonstrate the AmSI with selected axisymmetric simulations using the performance-portable Godunov code \textsc{idefix} \citep{lesur23}. To solve Eqs. \ref{eq:rhog}---\ref{eq:vdust} and \ref{eq:ind}, we enable the code's MHD, dust, and shearing box modules. We choose piecewise linear reconstruction, the HLLD Riemann solver, and second-order Runge-Kutta time integration. Ambipolar diffusion is integrated using a Runge–Kutta–Legendre super-time-stepping scheme. We model a single dust species with a constant stopping time\footnote{This is equivalent to a constant Stokes number in the shearing box.} and include feedback onto the gas. 

Our prescription for the ambipolar diffusivity in \textsc{idefix} follows the standard form (e.g. Cui \& Lin, 2021): 
\begin{align}
    \left.\frac{\p\bmB}{\p t}\right|_\mathrm{AD, \textsc{idefix}} = \nabla\times\left[\frac{\eta_\mathrm{AD}^\prime}{|\bmB|^2}\left(\bm{J}\times\bmB\right)\times\bmB \right],
\end{align}
with a constant $\eta_\mathrm{AD}^\prime$. This is related to the ambipolar diffusivity used above by $\eta_\mathrm{AD} = \eta_\mathrm{AD}^\prime B_z^2/|\bmB|^2$. At the linear level, this is only a constant factor and therefore has no impact on the dusty gas' stability properties. In Appendix \ref{code_test}, we present code tests against the AmSI linear modes. 

For the illustrative cases here, we use $\eta = 0.05h$ for the radial pressure gradient, and initialize the magnetic field with $\beta_z=10^4$, $B_y = -B_z$, and Elsassar number  $\Lambda_\mathrm{AD}^\prime \equiv c_{A,z}^2/\eta_\mathrm{AD}^\prime\Omega = 10$. For the dust, we consider an initial $\epsilon=0.2$ and $\st=0.01$ or $\st=0.1$. The initial dust and gas densities are uniform, with velocities given by Eqs. \ref{eq:uxgas}---\ref{eq:uydust}.

To isolate the AmSI, which have short characteristic lengthscales, we adopt a small computational domain with $x, z \in[-0.005,0.005]H_g$ to exclude MRI-related modes. We apply periodic boundaries in $z$ and shear-periodic boundaries in $x$, and use a resolution of $N_{x,z} = 512$. Note that the simulations are 3D but axisymmetric, which is realized by setting the azimuthal domain to one cell. To trigger instability, we add random perturbations to the gas radial velocities of amplitude $10^{-3}c_s$. Computational units are such that $H_g=\Omega=\mu_0=1$ and the initial $\rhog=1$.  

\begin{figure*}
    \centering
    \includegraphics[width=0.5\linewidth]{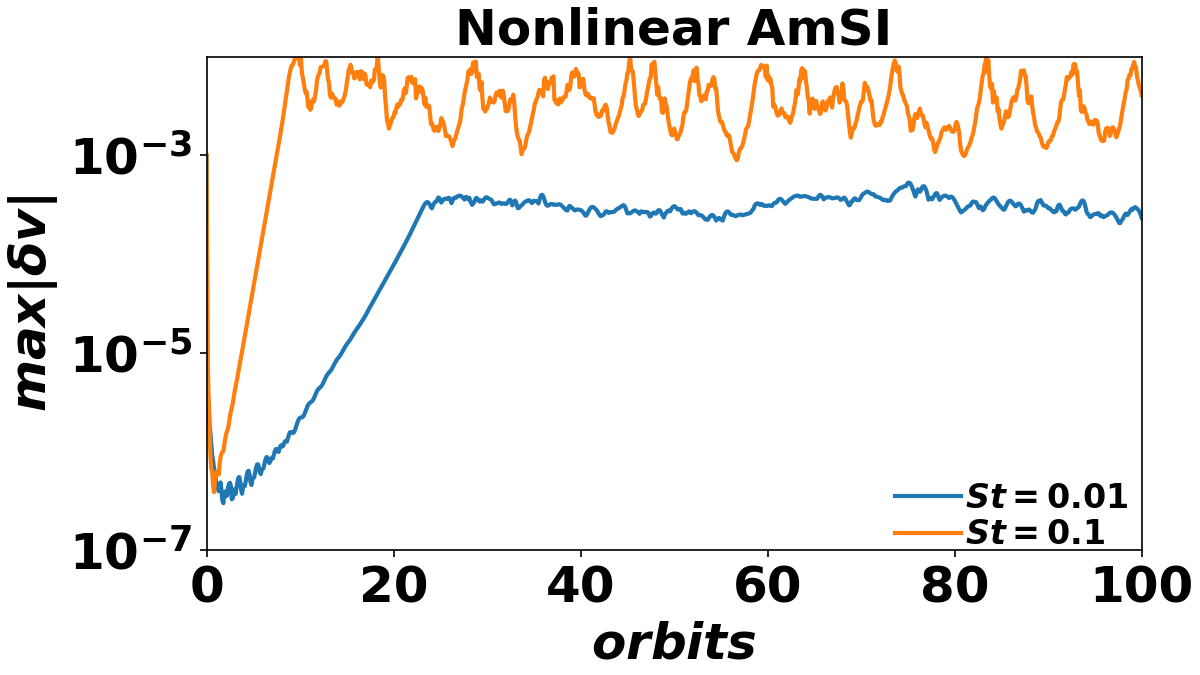}\includegraphics[width=0.5\linewidth]{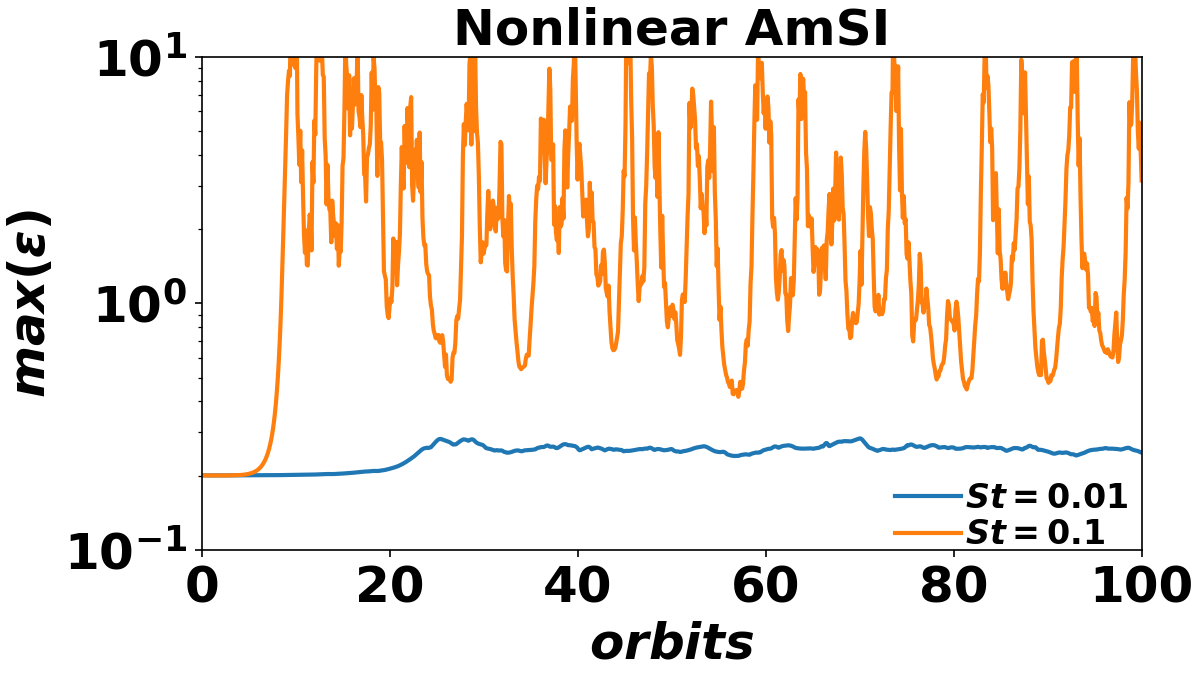}
    \caption{Nonlinear evolution of the AmSI with $\st=0.01$ (blue) and $\st=0.1$ (orange). Left: the maximum gas velocity perturbation. Right: the maximum dust-to-gas ratio.}
    \label{fig:amsi_evol}
\end{figure*}

\begin{figure*}
    \centering
    \includegraphics[width=0.5\linewidth]{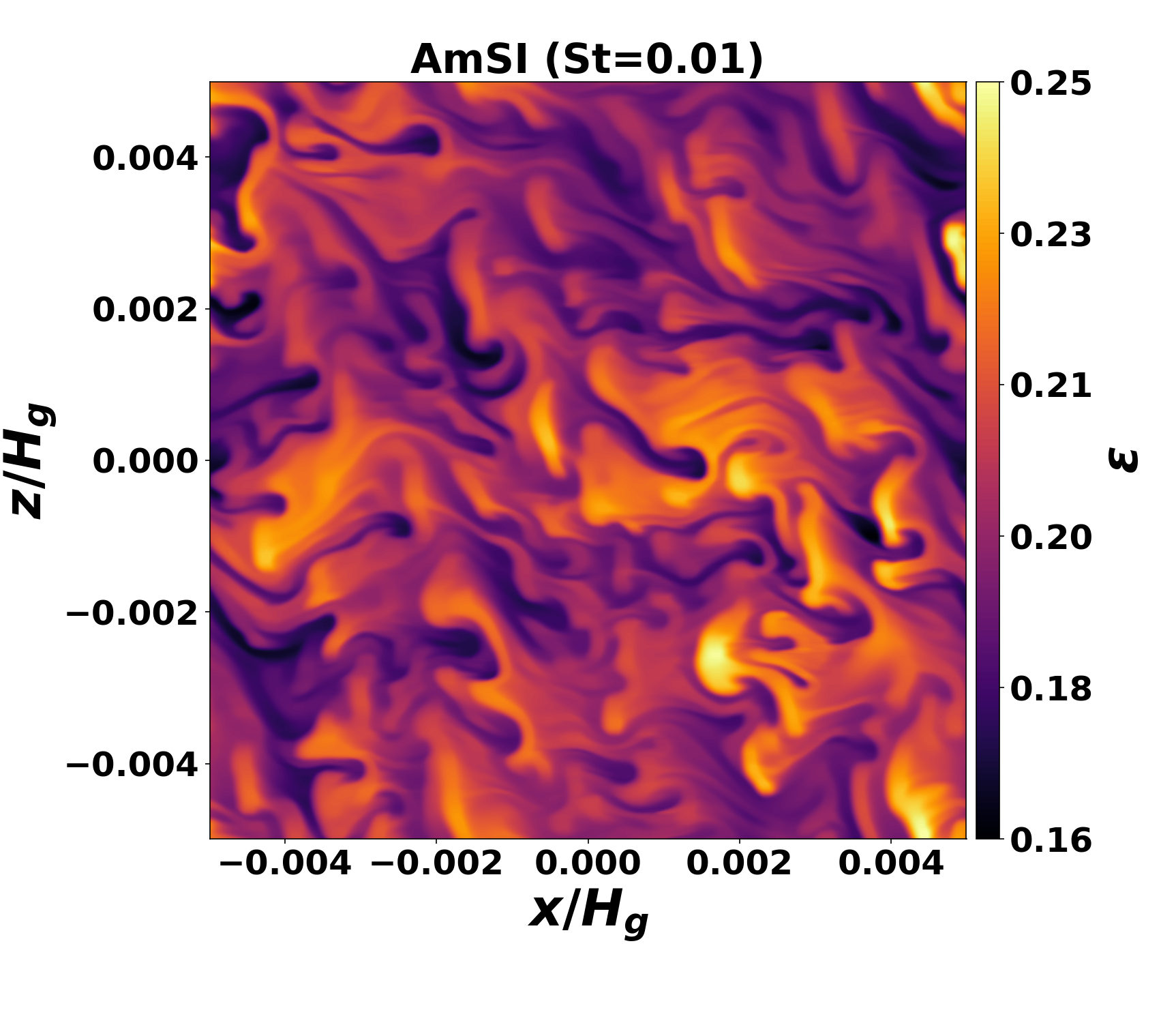}\includegraphics[width=0.5\linewidth]{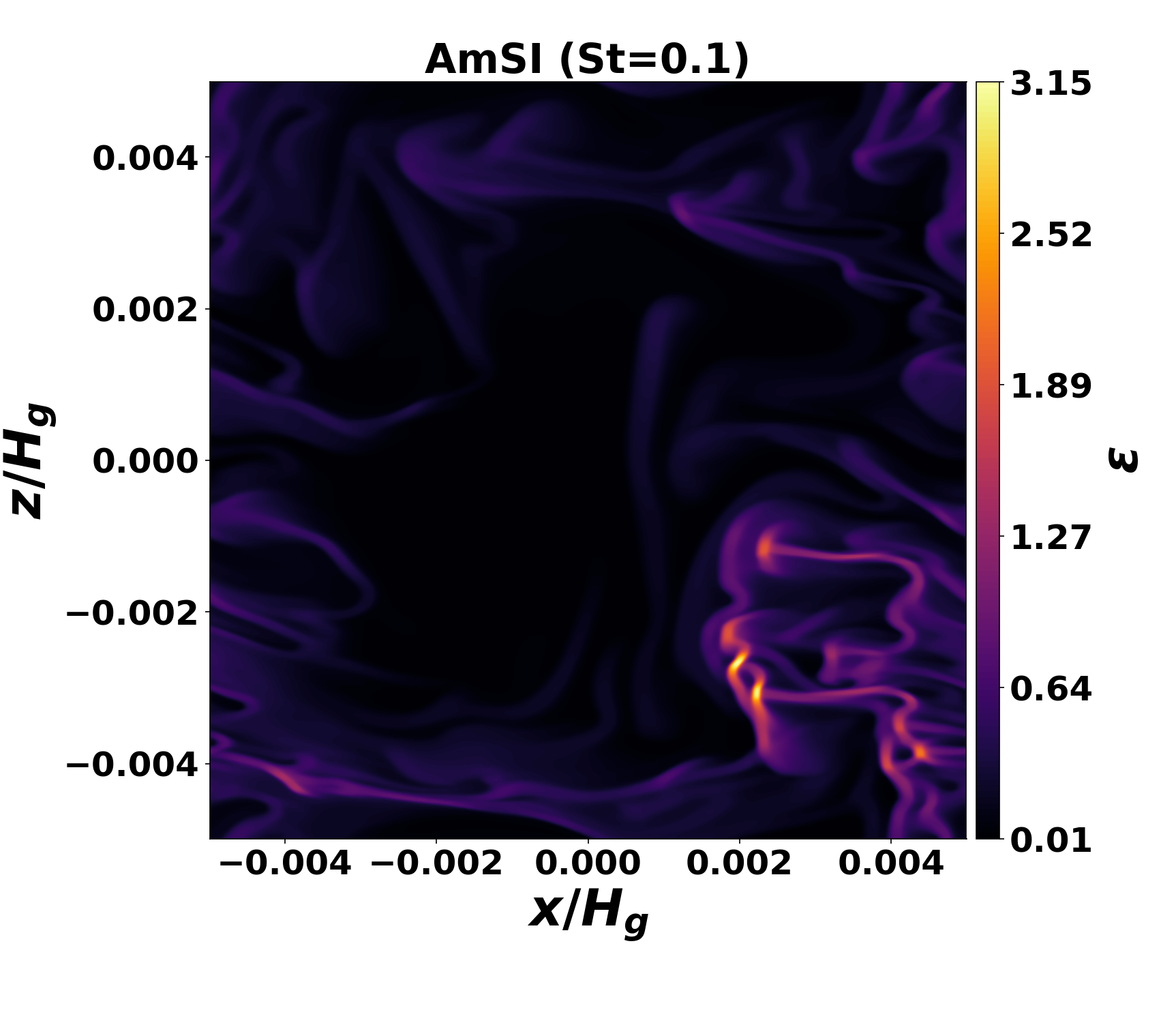}
    \caption{Snapshots of the dust-to-gas ratios  at the end of the AmSI simulations shown in Fig. \ref{fig:amsi_evol}. Left: $\st=0.01$. Right: $\st=0.1$. }
    \label{fig:amsi_snaps}
\end{figure*}

Fig. \ref{fig:amsi_evol} shows the evolution of the maximum gas velocity perturbations and dust-to-gas ratios for the two simulations; while Fig. \ref{fig:amsi_snaps} show snapshots of the dust-to-gas ratio. As expected, the systems are linearly unstable, with growth rates $\simeq 0.05\Omega, \, 0.2\Omega$ and saturation amplitudes $\operatorname{max}|\delta v|\sim 3\times 10^{-4}c_s, \, 3\times10^{-3}c_s$ for $\st=0.01$ and $\st=0.1$, respectively. We find for $\st=0.01$ dust concentrations are weak with $\operatorname{max}(\epsilon)\lesssim 0.3$, or about a $50\%$ increase from the equilibrium value.

On the other hand, for $\st=0.1$ we find dust densities can reach $\operatorname{max}(\epsilon)\sim 10$. The system undergoes strong oscillations, with rapid formation and destruction of transient clumps with $\epsilon\gtrsim 1$. See the right panel of Fig. \ref{fig:amsi_snaps} for an example clump. We find the average  $\operatorname{max}(\epsilon)\simeq 2$, i.e., 10  times  the initial value. This run can be compared with the classical SI case of \citet{johansen07a} that also used $\epsilon=0.2$ initially and $\st=0.1$ (their `AA' run), which found much  smaller dust overdensities at about $25\%$. This suggests that AmSI is more effective at concentrating dust in magnetized disks than in unmagnetized disks. However, it should be noted that MRI modes are absent by construction. The role of the AmSI in the presence of the MRI needs to be clarified in the future. 

\section{Discussion and conclusion}
\label{sec:sec7}
In this paper, we examined the effect of ambipolar diffusion (AD) on the stability of a dusty, magnetized disc.  It is expected the ionization fraction to be relatively low in the main body of the disc, such that ambipolar diffusion is believed to play an important role in the dynamics and evolution of protoplanetary discs. In particular, it has been shown that the combined effect of ambipolar diffusion and Ohmic resistivity can launch magnetically driven disc winds \citep{bai13, gressel15, bethune17} that drive accretion through  the disc surfaces. 

 We find that  dust feedback tends to stabilize the oblique modes of the MRI identified by \citet{kunz04}. This occurs as a result of a decrease in the gas azimuthal velocity, such that the feedback loop between ambipolar diffusion and differential rotation is weakened.   Depending on the orientation of the azimuthal field, our linear stability analysis also shows that ambipolar diffusion can give rise to a vigorous RDI,  resulting from the destabilization of  an Alfv\'en wave by dust-gas drift. The main effect of   ambipolar diffusion is to strongly modify  the Alfv\'en wave frequency, with the consequence that the range of wavenumbers exhibiting significant growth rates $\sigma\gtrsim  0.1 \Omega^{-1}$ is significantly extended in comparison with a standard Alfv\'en wave RDI. 

  Our analysis shows that  ambipolar diffusion  has a stronger impact when the azimuthal field amplitude $B_y$ is comparable to the vertical field $B_z$, and for ambipolar Elsasser numbers $\Lambdaad\sim 10$ typically. In the outer regions of protoplanetary discs (PPDs) where AD is the  dominant non-ideal MHD effect,  $\Lambdaad \sim 1$ in the disc midplane \citep{bai2011}, while  amplitudes of azimuthal fields such that $B_y/B_z=\mathcal{O}(10)$ have been found in calculations incorporating AD \citep{cui21}. Under these conditions, the requirements for efficient instability would generally not be satisfied.  At smaller radii $R\sim 1-20$ AU, however, global simulations of magnetized disc winds show that the toroidal and vertical fields can have comparable amplitudes \citep{bai2017}, particularly at heights $Z\lesssim 2H$, with $H$ the gas scale height \citep{gressel15}. In these inner disc regions, $\Lambdaad$ is still expected to be of order unity but can also be significantly higher in the presence of large grains. For example, at distances $R\sim 10$ AU,  \citet{simon2015}  reported a value  $\Lambdaad\sim 10$ at the midplane of a disc that containing  $10\; \mu m$ dust grains. These results therefore suggest that the instability is most likely to operate in the intermediate regions of wind launching discs, particularly where a substantial population of large grains is present. 
 
 Interestingly, we obtain significant growth rates in dust-poor discs with $\epsilon\sim 0.2$ and for Stokes numbers as low as $\st=10^{-4}$, although numerical simulations suggest that dust enhancements resulting from unstable AmSI modes might be relatively modest in that case. For higher Stokes numbers with $\st=0.1$, however, we obtain more significant dust concentrations compared to an unmagnetized disc, which  may facilitate planetesimal formation in magnetized discs subject to ambipolar diffusion. 

  A main limitation of this work is that the effect of vertical stratification has not been considered. In the hydrodynamical limit and for $\st \sim 0.1$, \citet{lim2025} recently obtained good agreement between stratified and unstratified simulations of the streaming instability, such that the midplane dynamics can be well described using an unstratified setup. In a magnetized disc launching a magnetocentrifugal wind, however, AD tends to  generate a flip of the toroidal field close to the disc midplane \citep{bethune17}. In this context, numerical simulations investigating the effect of AD on the streaming instability should ideally take into account the vertical  structure of the disc. Numerical simulations of vertically global, radially unstratified wind-emitting dusty discs will be presented in a future study.

\begin{acknowledgements}
Computer time for this study was provided by the computing facilities MCIA (M\'esocentre de Calcul Intensif Aquitain) of the Universite de Bordeaux and by HPC resources of Cines under the allocation A0170406957 made by GENCI (Grand Equipement National de Calcul Intensif). Simulations were also carried out on the \emph{Kawas} cluster at ASIAA. The authors acknowledge the access to high-performance computing facilities (Theory cluster and storage) provided by ASIAA. MKL is supported by the National Science and Technology Council (grants 113-2112-M-001-036-, 114-2112-M-001-018-, 113-2124-M-002-003-, 114-2124-M-002-003-, 115-2124-M-002-014-), an Academia Sinica Career Development Award (AS-CDA-110-M06), and an Academia Sinica Grand Challenge Seed Grant (AS-GCS-115-M02).
\end{acknowledgements}

\bibliographystyle{aa}
\bibliography{ref}

\begin{appendix}

\section{Single-fluid Model}
\label{sec:appA}
In order to get insight of the effect of dust on the on the oblique modes of \cite{kunz04}, we employ a more simple model of a dusty,  magnetized gas. In this approach, we define the total dust density:
\begin{equation}
\rho=\rho_g+\rho_d
\end{equation} 
and the center of mass velocity:
\begin{equation}
\utotal=f_g\ugas+f_d\vdust
\end{equation} 
where $f_g=1/(1+\epsilon)$ is the gas fraction and $f_d=\epsilon/(1+\epsilon)$ the dust fraction.  We work within the framework of the magnetized terminal velocity approximation, which corresponds to a first order approximation in terms of the Stokes number, and which is therefore better suited to small particles. In this regime, the velocity difference between the dust and gas phases $\Delta \utotal=\vdust-\ugas$ is given by:
\begin{equation}
\Delta \utotal=\tau_s\left[\frac{\nabla P}{\rho}-2\eta R\Omega^2f_g\ex -\frac{1}{\mu_0 \rho} (\nabla\times \bfield)\times \bfield \right]
\label{eq:tva}
\end{equation}
In this limit, the gas incompressibility condition, dust continuity equation, and the center-of-mass momentum equation for the dust-gas mixture are:
\begin{equation}
\nabla \cdot \utotal=\nabla \cdot (f_d \Delta \utotal)
\end{equation}
\begin{equation}
\frac{\partial \epsilon}{\partial t}+\nabla \cdot (\epsilon  \utotal)=-\nabla \cdot  \utotal
\end{equation}
\begin{equation}
\frac{\partial \utotal}{\partial t}+\utotal\cdot \nabla  \utotal=-\frac{1}{\rho} \nabla P+2\eta r \Omega^2 f_g \ex+2 \Omega u_y \ex-\frac{\Omega}{2}u_x \ey+\frac{1}{\mu_0 \rho}(\nabla \times \bfield)\times \bfield
\end{equation}

In principle, the original induction equation given in Eq.  \ref{eq:ind} should be expressed in terms of $\utotal$ and $\Delta \utotal$. Following \cite{lin22}, we do not take this into account  and simply set $\ugas\rightarrow \utotal$ in Eq.  \ref{eq:ind}, such that the induction equation is leaved unchanged in the single fluid model. We revisit this approximation in \S\ref{dusty_am}.

\subsection{Linearized equations}
\begin{equation}
\nabla \cdot \delta \utotal=-\frac{\sigma}{1+\epsilon}\delta \epsilon
\end{equation}
\begin{equation}
\begin{aligned}
\sigma\delta \epsilon=
&2ik_x\tau_s \eta R \Omega^2 f_g\frac{1-\epsilon}{1+\epsilon} \delta \epsilon+f_d\tau_s k^2 (1+\epsilon)\frac{\delta P}{\rho}\\
&+f_d\tau_s\frac{B_z}{\rho \mu_0}k^2\left(-\frac{k_x}{k_z}\delta B_x +\frac{B_y}{B_z}\delta B_y\right)
\end{aligned}
\end{equation}
\begin{equation}
\sigma \delta u_x=2\Omega \delta u_y-ik_x\frac{\delta P}{\rho}+i\frac{k^2}{k_z}\frac{B_z}{\rho \mu_0}\delta B_x-ik_x \frac{B_y}{\rho \mu_0}\delta B_y-\frac{2\eta R \Omega^2 f_g}{(1+\epsilon)^2}\delta \epsilon  
\end{equation}
\begin{equation}
\sigma \delta u_y=-\frac{\Omega}{2} \delta u_x+ik_z \frac{B_z}{\rho \mu_0}\delta B_y
\end{equation}
\begin{equation}
\sigma \delta u_z=-ik_z\frac{\delta P}{\rho}-ik_z \frac{B_y}{\rho \mu_0}\delta B_y
\end{equation}
\begin{equation}
\sigma \delta B_x=ik_z B_z \delta u_x+\frac{1}{\gamma \rho_i \rho_g}(-k^2B_z^2\delta B_x+k_xk_zB_z B_y \delta B_y)
\end{equation}
\begin{equation}
\begin{aligned}
\sigma \delta B_y=
&ik_z B_z \delta u_y-i(k_x\delta u_x+k_z\delta u_z)B_y-\frac{3}{2}\Omega\delta B_x \\
&+\frac{1}{\gamma \rho_i \rho_g}(k^2\frac{k_x}{k_z}B_zB_y\delta B_x-(k^2B_y^2+k_z^2B_z^2)\delta B_y)
\end{aligned}
\end{equation}
\subsection{Dispersion relation}
 We find that the dispersion relation can be written under the form:
 \begin{equation}
 \begin{split}
 (f_d\sigma^6+\dcoef_5\sigma^5+\dcoef_4\sigma^4+\dcoef_3\sigma^3+\dcoef_2\sigma^2+\dcoef_1\sigma+\dcoef_0)\tau_s\\
 +(\sigma^4+\ccoef_3\sigma^3+\ccoef_2\sigma^2+\ccoef_1\sigma+\ccoef_0) \sigma=0
 \label{eq:disperse}
 \end{split}
 \end{equation}
 where the coefficients $\ccoef_i$ are those given in Eqs. 32-35 of \cite{kunz04} and the coefficients $\dcoef_i$ are defined by:
 \begin{equation}
 \dcoef_5=\frac{f_d}{\gamma \rho_i \rho_g}\left(k^2B^2+k_z^2B_z^2\right)
 \end{equation}
 \begin{equation}
 \begin{aligned}
 \dcoef_4=&2f_d\caz^2k_z^2+\Omega^2(f_d-2i\eta Rk_x)+\frac{3}{2}f_d\frac{B_zB_y}{\gamma \rho_i \rho_g}k_xk_z\Omega\\
 &+f_d\left(\frac{BB_z}{\gamma \rho_i \rho_g}\right)^2k^2k_z^2
 \end{aligned}
 \end{equation}
\begin{equation}
 \begin{aligned}
\dcoef_3=&-2f_d\frac{B_y}{B_z}(\caz k_z)^2\frac{k_x}{k_z}\Omega\\
&+\frac{B^2k^2+B_z^2k_z^2}{\gamma \rho_i \rho_g}\left(f_d(\caz k_z)^2+\Omega^2(f_d-2i\eta Rk_x)\right)
 \end{aligned}
\end{equation}
\begin{equation}
 \begin{aligned}
 \dcoef_2=&f_d(\caz k_z)^4-(\caz k_z)^2(3f_d+4i \eta R k_x)\Omega^2-2i \eta \frac{k_xk_z^2}{k^2}\Omega^4\\
 &+\frac{B_zB_y}{2\gamma \rho_i \rho_g}k_xk_z
 \left(3(f_d-2i\eta Rk_x)\Omega^2-f_d(\caz k_z)^2 \right)\\
 &+\left(\frac{BB_z}{\gamma \rho_i \rho_g}\right)^2k^2k_z^2(f_d-2i\eta Rk_x)\Omega^2
  \end{aligned}
\end{equation}
\begin{equation}
 \begin{aligned}
 \dcoef_1=&-2f_d\frac{B_y}{B_z}\frac{k_x}{k_z}(\caz k_z)^4\Omega\\
& +\frac{B_z^2}{\gamma \rho_i \rho_g}\left(\frac{B_y^2}{B_z^2}\left[(\caz k_x)^2(f_d-2i\eta Rk_x)-2i \eta k_x (\caz k_z)^2\right]k_z^2 \Omega^2 \right.\\
 &\left.-2i\eta R k_x\left(1+\frac{k_z^2}{k^2}\right)(\caz^2k^2+\Omega^2)\right)
 \end{aligned}
 \label{eq:d1}
 \end{equation}
\begin{equation}
 \begin{aligned}
  \dcoef_0=&-2i(\caz k_z)^2 \eta R k_x\left((\caz k_z)^2-3\Omega^2\frac{k_z^2}{k^2}\right)\Omega^2\\
  &-3i\frac{B_yB_z}{\gamma \rho_i \rho_g}\eta R k_x^2k_z\left((\caz k_z)^2+\Omega^2\frac{k_z^2}{k^2}\right)\\
  &-2i \left(\frac{BB_z}{\gamma \rho_i \rho_g}\right)^2\eta R k_x k_z^4 \Omega^4
   \end{aligned}
  \end{equation}
  \subsection{Limit of strong non-ideal effects}
  In the limit where ambipolar diffusion is very strong, namely for Elsasser numbers $\Lambda_A \ll 1$, an approximation to the dispersion relation can be obtained by retaining only terms 
  $\propto \Lambda_A^{-2}$. In this regime, we find that the dispersion relation becomes:
  \begin{equation}
  f_g \tau_s\sigma^4+\sigma^3+\tau_s(f_d\Omega^2-2i \eta k_x R \Omega^2)+\frac{k_z^2}{k^2}\Omega^2 \sigma-2ik_x\tau_s\eta R\Omega^4\frac{k_z^2}{k^2}=0
  \end{equation}
  which is equivalent to classical SI relation dispersion \citep{jacquet11, laibe14, lin17}. For $\st \ll 1$ and $\sigma={\cal O}(\st)$, the first term in the previous expression can be neglected, and we recover the cubic dispersion relation of \cite{youdin05}.

\subsection{Effective ambipolar diffusion from dust-gas drag}\label{dusty_am}
Recall that we neglected $\Delta\utotal$ in the induction equation for the above analysis. To evaluate the consequence of this approximation, we note that:
\begin{equation}
\ugas=\utotal-f_d \Delta \utotal
\end{equation}
It follows that the induction equation, in the ideal MHD limit,  can be rewritten as:
\begin{equation}
\frac{\partial \bfield}{\partial t}=\nabla\times (\utotal \times \bfield)
- \frac{3}{2}\Omega B_x \ey
-\nabla\times (f_d \Delta \utotal \times \bfield)
\label{eq:ind2}
\end{equation}
Using  the magnetized terminal velocity approximation given by Eq. \ref{eq:tva}, the previous expression becomes:
\begin{align}
\frac{\partial \bfield}{\partial t}=&\nabla\times (\utotal \times \bfield)- \frac{3}{2}\Omega B_x \ey 
-\nabla\times\left[f_d\tau_s\left(\frac{\nabla P}{\rho} - 2\eta R\Omega^2f_g\ex\right)\times\bfield\right]\notag\\
&+\nabla\times\left\{\tau_s f_d f_g c_A^2\left[(\nabla \times \bfield)\times \eb\right] \times \eb \right\}
\label{eq:ind2}
\end{align}
with $c_A^2=B^2/\mu_0 \rho_g$  and $\eb=\bfield/B$. 
Therefore,  the relative drift between dust and gas phases has the same effect of the drift between neutral and charged species in MHD, as it also leads to ambipolar-like diffusion in the final term on the right hand side.

From the previous equation, the dust effective ambipolar resistivity would be given by $\eta_{A, {\rm eff}}=\tau_s f_d f_g c_A^2$, corresponding to an effective ambipolar Elsasser number:
\begin{equation}
\Lambda_{A, {\rm eff}}=\frac{1}{\st f_d f_g}.
\end{equation} 
For small grains with $\st\ll 1$, we have $\Lambda_{A, {\rm eff}}\gg1$ and thus this effect is likely unimportant compared to the true ambipolar diffusion in protoplanetary discs.

\section{Code test}\label{code_test}

\begin{figure}
    \centering
    \includegraphics[width=\linewidth]{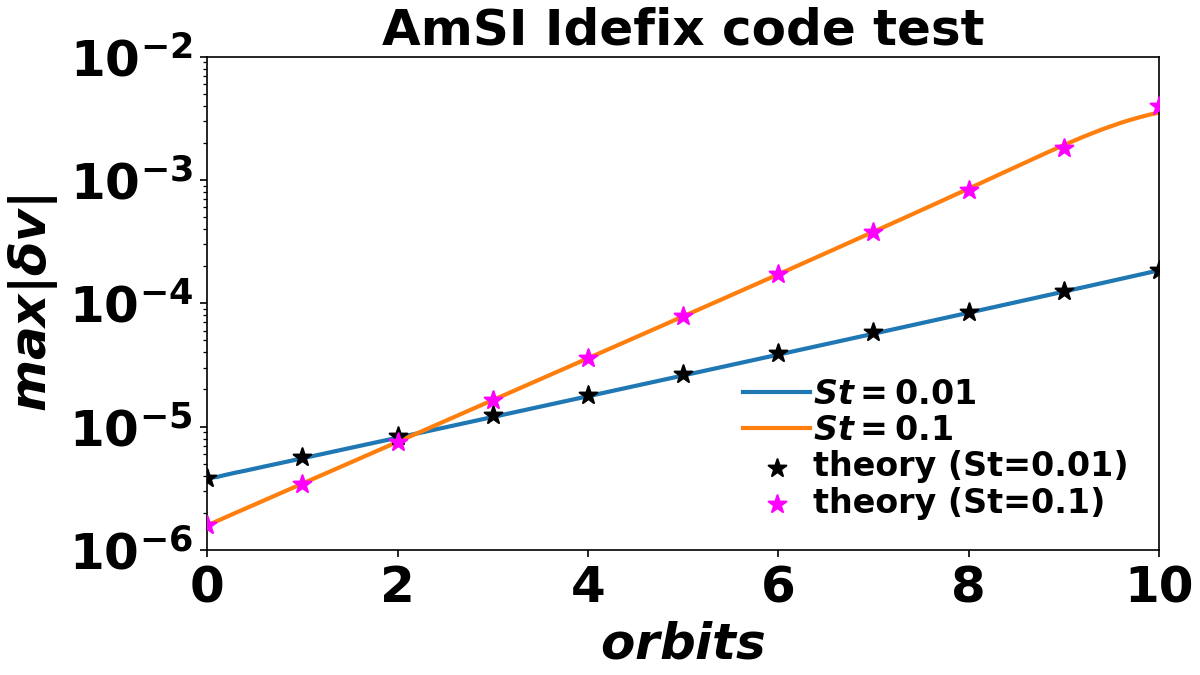}
    \caption{Evolution of the maximum gas velocity perturbations for \textsc{idefix} runs initialized with clean eigenmodes (blue and orange curves). The stars are the expected evolution based on linear theory growth rates. 
    \label{fig:amsi_code_test}}
\end{figure}

We test our \textsc{idefix} setup by simulating a clean eigenmode and comparing its growth rate against linear theory. The equilibrium disk parameters are the same as that used for Sect. \ref{sec:sec6}. We perturb the equilibrium by adding axisymmetric perturbations calculated from linear theory (i.e. Eqs. \ref{eqi}-\ref{eqf} with $\etaad$ replaced by $\eta_\mathrm{AD}^\prime$), which are characterized by wavenumbers $k_{x,z}$. We normalize the perturbations so that $\operatorname{max}|\delta v_x|=10^{-6}c_s$. We consider two cases: \begin{inparaenum}[1)]
\item $\st=0.01$ with $k_{x}H_g = 4000$, $k_zH_g=1500$; and
\item $\st=0.1$ with $k_{x,z}H_g = 1000 $. 
\end{inparaenum}
For these tests, the computational domain is one wavelength, i.e. $x, z\in [-\pi/k_{x,z}, \pi/k_{x,z}]$, and we use $N_{x,z}=128$ cells in each direction. 

Fig. \ref{fig:amsi_code_test} shows the evolution of the maximum gas velocity perturbations, from which we measure growth rates over the first 8 orbits and also compare them in Table \ref{tab:eigenmode-test}. These show a good agreement between simulation and theoretical growth rates, with a relative error $\lesssim 1\%$. Not surprisingly, the $\st=0.1$ case, with a higher growth rate, is more accurately simulated.

\begin{table*}[ht]
\centering
\caption{Comparison of growth rates, $\Re{(\sigma)} / \Omega$, between \textsc{idefix} simulations initialized with clean eigenmodes and linear theory predictions.} 
\label{tab:eigenmode-test}
\begin{tabular}{lcccc}
\toprule
St & $k_{x,z}H_g$ & $\Re{(\sigma}) / \Omega$ (\textsc{idefix}) & $\Re{(\sigma)} / \Omega$ (theory) & Rel. error (\%) \\
\midrule
0.01 & 4000, 1500         & \num{6.170951161677132e-02} & \num{6.179272480815939e-02} & -0.13 \\
0.10 & 1000, 1000         & \num{1.248219723748917e-01} & \num{1.247635506629079e-01} & +0.05 \\
\bottomrule
\end{tabular}
\end{table*}

The test here shows that \textsc{idefix} is capable of capturing the AmSI. We have also reproduced growth rates with $N_{x,z}=64$ cells at similar accuracy by using a less diffusive configuration: piecewise parabolic reconstruction, a Roe solver, and third order Runge-Kutta time integration. However, we found this setup is prone to crashing (with vanishing timesteps) when evolving the AmSI into the nonlinear regime. Hence we do not use it in Sect. \ref{sec:sec6}.

\end{appendix}

\end{document}